\documentclass[a4paper,12pt]{article}
\usepackage{amssymb,amsmath,mathbbol,mathrsfs}
\usepackage[usenames,dvipsnames]{color}
\usepackage{hyperref}
\usepackage{xcolor}
\usepackage{stmaryrd}
\usepackage{authblk}
\usepackage{framed}
\usepackage{empheq} 
\usepackage{slashed}
\usepackage{hyphenat}
\usepackage{apacite}
\usepackage{natbib}
\usepackage{dsfont}
\usepackage{chngcntr}
\usepackage{enumerate}  
\usepackage{tikz-cd}
\usepackage{booktabs, array, multirow, makecell}

\newcolumntype{C}[1]{>{\centering\arraybackslash}p{#1}}

\usepackage{marginnote}    

\usepackage[left=.8in,right=.8in,top=.8in,bottom=.8in]{geometry}                  % For good copy.

\usepackage{sectsty} 
\allsectionsfont{\sffamily\mdseries\upshape} % (See the fntguide.pdf for font help)
\usepackage{tocloft}

\makeatletter
\renewenvironment{abstract}{%
    \if@twocolumn
      \section*{\abstractname}%
    \else %% <- here I've removed \small
      \begin{center}%
        {\bfseries\sffamily\abstractname\vspace{\z@}}%  %% <- here I've added \Large
      \end{center}%
      \quotation
    \fi}
    {\if@twocolumn\else\endquotation\fi}
\makeatother

\numberwithin{equation}{section}
\hypersetup{
	colorlinks=true,         
	linkcolor=MidnightBlue,          
	citecolor=BrickRed,        
	urlcolor=MidnightBlue            
}

\newcommand{\BIBand}{and}%{\&}

\newcommand{\be}{\begin{equation}}
\newcommand{\ee}{\end{equation}}

\renewcommand{\d}{{\mathrm{d}}}

\newcommand{\Ad}{{\mathrm{Ad}}}

\renewcommand{\bar}{\overline}

\newtheorem{lem}{Lemma}
\newcommand{\cint}{{\int\kern-.87em{<}}}
\newcommand{\sint}{{\int\kern-.75em{\sim}}}
\newcommand{\fint}{{\int\kern-1.00em{\int}}}

\newcommand{\bb}{\mathbb}

\newcommand{\order}[1]{\ensuremath{\mathcal{O}(#1)}}

\renewcommand{\#}{\sharp}
\newtheorem{prop}{Proposition}
\newtheorem{proof}{Proof}

\let\oldmarginpar\marginpar
\renewcommand\marginpar[1]{\oldmarginpar{\color{red}\raggedright\footnotesize #1}}

\begin{document}

\title{Charge quantisation without compactness: the Higgs and Yukawa mechanisms from matter geometry}
\author{Henrique Gomes\\
\normalsize{Oriel College, University of Oxford, and Lichtenberg Center, University of Bonn.}}

\maketitle

\begin{abstract}
In the \emph{geometry-first} formulation of gauge theory
\citep{Gomes_internal}, Hermitian vector bundles and their covariant
derivatives are the primitive objects. Here gauge groups arise only as the
automorphism groups of a small set of structured fundamental fibres, and
every matter field is a section of a tensor product of the fundamental
bundles. This is a constraint on gauge theories. In the abelian sector it
enforces charge quantisation, even without
compactness: charges count tensor powers of a fundamental line bundle, so
they lie on an integral lattice even when the automorphism group is the
non-compact $\mathbb{C}^\times$, whereas a symmetry-first theory built on
$\mathbb{R}$ admits irrationally related charges. The Standard Model
satisfies the constraint, and its two mass-generating mechanisms follow
from the fibre geometry. The Higgs mechanism arises from the
second fundamental form of the sub-bundle singled out by the Higgs vacuum. This works with no appeal to
symmetry breaking or Goldstone's theorem, and the full electroweak
spectrum, $m_W=m_Z\cos\theta_W$ included, is computed from the same
operator. Each Yukawa coupling is the
canonical fibrewise contraction fixed by the inner products and volume
forms of the fundamental bundles. Exceptional gauge groups remain formally presentable, but on
progressively less elementary fibre structure; at $E_8$ the fundamental
fibre is the Lie algebra $\mathfrak{e}_8$ itself, and the claim of conceptual priority collapses. A companion paper shows the two formulations are not
equivalent, in a categorical sense \citep{Gomes_nonequiv}.
\end{abstract}

\tableofcontents
\section{Introduction}\label{sec:intro}

In the Standard Model the left- and right-handed components of a massive fermion carry inequivalent gauge representations. Therefore they cannot form a bare Dirac mass term: there is no non-zero invariant bilinear pairing between the internal spaces of $\psi_L$ and $\psi_R$. The Higgs is what allows a mass term by allowing an invariant trilinear
\be\label{eq:trilinear_intro}
\bar V_L\otimes W\otimes V_R\longrightarrow\mathbb{C},
\ee
with $W$ the Higgs representation or its conjugate. The space of such trilinears is one-dimensional.

In this case, representation theory says nothing about where the contraction comes from. The geometry-first formulation must get it from the fibre geometry, and indeed, that geometry supplies a preferred element of the line. The Hermitian metrics yield the charged-lepton and down-type contractions, and the up-type contraction uses in addition the unit complex volume form on the weak fibre. The scalar couplings, and with three generations the Yukawa matrices, remain free.

The same question can be put to the rest of the Standard Model's structure. How much of it follows from vector bundles with a certain fibre geometry and compatible covariant derivatives, with the gauge groups recovered as automorphism groups?

There is clear precedent for the question. Symmetry principles have been among the most reliable guides to the laws of physics. They are also empirically accessible in a  straightforward way. One can discover that distinct particles scatter alike, or that spectra exhibit degeneracies, well before one has a dynamical theory to explain those patterns (cf.\ \citealp{Brading_Castellani}; \citealp{Lange_metalaws}). In particle physics this empirical route (particle $X$ behaves like particle $Y$ under condition $A$, so postulate a symmetry relating them) has been the standard path to the gauge groups of the Standard Model.

However, symmetries discovered in this way often have a characteristic subsequent career. They are first treated as constraints on the laws, but are often then promoted into features of a geometric stage on which the physics plays out. Lorentz covariance, read off from electrodynamics, was absorbed into Minkowski geometry; Einstein's principle of general covariance was successfully absorbed into pseudo-Riemannian geometry; even Galilean symmetry can be absorbed by Newton-Cartan geometry. Once the promotion is complete, the symmetry that had seemed foundational turns out to be derivative: a consequence of the geometry rather than an independent postulate.\footnote{This pattern is contested: \citet{Brown_PR} argues that the geometric formulation can obscure the dynamical reasoning that led to the symmetry. I take his point, but the consolidation remains valuable even when the dynamical route retains independent interest.}

Gauge theory has not yet matured in the same way. The standard mathematical formulation, using principal fibre bundles, is geometric, to be sure: bundles and connections are the geometer's stock-in-trade. But it also posits the structure group $G$ as an independent ingredient that coordinates matter sectors `from outside'. The group is not read off from the spaces where matter fields live: it is postulated independently. I call this a \emph{symmetry-first} approach to gauge theory.

Earlier work (\citealp{Gomes_internal, Gomes_AB}) introduced a \emph{geometry-first} alternative that aims to carry the career of gauge symmetry to completion, at least for a significant class of theories. The fundamental objects are then vector bundles equipped with compatible covariant derivatives; the group $G$ appears only as the automorphism group of the fibre, and principal bundles play no foundational role. For the Standard Model, three fundamental Hermitian vector bundles (with fibres $\mathbb{C}^3$, $\mathbb{C}^2$, and $\mathbb{C}$) suffice: matter fields are sections of their tensor products, and the gauge group $SU(3)\times SU(2)\times U(1)$ emerges as a product of fibre-automorphism groups. Throughout, what is eliminated is only the gauge group as an independent posit.

It bears mentioning that the ambition of recovering the Standard Model's gauge group and representation content from prior structure is shared by two other families of programmes. In the spectral programme of Connes and collaborators \citep{ChamseddineConnes_spectral,CCM_gravity}, the input is a finite noncommutative algebra with a Dirac operator.%: the gauge group arises from the algebra's unitaries, the Higgs from the finite part of a generalised connection, and the Yukawa matrices are absorbed into the geometry as components of the finite Dirac operator. The present framework geometrises less: the contraction is canonical, and the couplings and Yukawa matrices remain free parameters (Section~\ref{sec:Yukawa}).
\footnote{The spectral programme also illustrates the empirical risk this genre can carry: with the `big desert' assumption it predicted a Higgs mass near $170$\,GeV, subsequently excluded; compatibility with the observed $125$\,GeV was restored by promoting to a field what the earlier treatment had fixed as a scalar \citep{ChamseddineConnes_resilience}. For a reformulation of the framework's axioms, see \citet{BoyleFarnsworth_NJP}.} A second family derives the group from a distinguished algebra: division-algebraic ladder operators \citep{Furey_ladder}, octonionic constructions \citep{Krasnov_SO9}, and the exceptional Jordan algebra $J_3(\mathbb{O})$ \citep{DVTodorov_Jordan,Boyle_triality,BaezSchwahn_Jordan}; \citet{BaezHuerta_GUT} survey the representation-theoretic background. 

The present programme differs from both of these  in what it takes as primitive and in what it seeks to  recover. %: the fundamental fibres carry only standard geometric structures---Hermitian metrics, volume forms---and each gauge factor must be the \emph{full} automorphism group of one of them. The algebraic programmes recover the group instead as the stabiliser of additional structure inside a larger automorphism group; 
The cost of that difference, by the present framework's lights, is assessed in Section~\ref{sec:defense}.

So much for the programme in outline. Whether the explicit appeal to symmetry can actually be dispensed with is still far from obvious; representations of Lie groups, Killing forms, Casimir invariants, stabilisers, spontaneous symmetry breaking---these are the daily bread of the Standard Model. This paper shows that two central mechanisms  of particle physics (the Higgs mechanism and the Yukawa coupling) can be rederived without any of this apparatus, and that the resulting picture ties the gauge group to matter geometry closer than the principal-bundle formulation requires. In more detail, the paper makes three contributions in this regard.

\emph{Higgs mechanism.} I derive the classical Higgs mechanism entirely in vector-bundle language (Section~\ref{sec:Higgs}). A non-vanishing Higgs section singles out a rank-one sub-bundle $L\subset E$; the covariant derivative of the unit Higgs section defines a second fundamental form, that I interpret as a shape operator (or extrinsic curvature) for the orthogonal complement $L^\perp$. The quadratic dependence on the connection that characterises mass acquisition arises from this shape operator. Components of the connection that annihilate the Higgs section remain massless---corresponding, in symmetry-first language, to the unbroken gauge group. Neither symmetry breaking nor Goldstone's theorem plays any role; a gauge-fixing is obviated by a choice of representational scheme that parametrises non-isomorphic Higgs configurations directly (Proposition~\ref{prop:param}), without introducing and then eliminating redundant modes. The construction reproduces the standard electroweak spectrum explicitly: $m_W=\tfrac12 g v_{\mathrm{EW}}$, $m_Z=\tfrac12\sqrt{g^2+g'^2}\,v_{\mathrm{EW}}$, $m_\gamma=0$, with $m_W=m_Z\cos\theta_W$ (Section~\ref{sec:SU2}).

\emph{Yukawa mechanism.} I reconstruct the Standard Model Yukawa forms as fibrewise contractions on the fundamental bundles $E^1, E^2, E^3$ (Section~\ref{sec:Yukawa}), with the one-dimensionality of the invariant spaces proved as Lemma~\ref{lem:mult_one}. A normaliser computation (Proposition~\ref{prop:central}) shows that the residual freedom in the symmetry-first formulation changes only the choice of generator of an invariant line and introduces no further physical freedom. The up-type Yukawa contraction depends essentially on the unit volume form of the weak fibre.%, whose stabiliser in $U(2)$ is $SU(2)$; in the product-group presentation used here, that is what distinguishes them.

\emph{Scope and constraints.} I examine the relationship between the two formulations (Section~\ref{sec:defense}). The geometry-first picture requires each factor of $G$ to arise as the automorphism group of a fundamental bundle: a tighter alignment between group and geometry than the principal-bundle picture demands. The Standard Model satisfies this condition with Hermitian metrics and volume forms. Exceptional groups are formally presentable as well, but don't strictly speaking fit the `geometry-first' explanatory schema. In increasing degree of `lack of fit', we have $G_2$ fits the best of the lot, since it arises from the automorphisms of a standard three-form; $E_6$, $F_4$ and $E_7$ require for this description tensors descended from the exceptional Jordan algebra; and $E_8$ requires $\mathfrak{e}_8$ and its bracket as structures of the fundamental vectorial fibre, at which point the group is being recovered from itself. Charge quantisation is also enforced by construction: hypercharge labels count tensor powers of the fundamental line bundle $E^1$ (a discrete structure persisting even when the automorphism group is non-compact). More generally, a companion paper \citep{Gomes_nonequiv} shows that the two formulations are not mathematically equivalent: a natural functor from geometry-first data to principal bundles with matter is faithful but neither essentially surjective nor full. A second companion \citep{Gomes_globalform} argues that,  for the global form of the Standard Model gauge group, the non-equivalence gives an abductive argument towards unification, even at the classical level. 

\emph{Relation to prior work.} The formulation itself is not new here: \citet{Gomes_internal} introduced it and argued that principal bundles are dispensable for matter sectors. \citet{Gomes_AB} applied it to the Aharonov--Bohm effect. But neither paper contains any treatment of the Standard Model's mass-generating mechanisms (the shape-operator derivation of the Higgs mechanism, the electroweak spectrum with the Weinberg relation, the canonical Yukawa contractions with the one-dimensionality of the invariant spaces (Lemma~\ref{lem:mult_one}), and the CKM matrix as an eigenframe mismatch). These appear here for the first time, as do the alignment condition \eqref{eq:aut} and the graded analysis of the exceptional groups (Section~\ref{sec:defense}). The two companion papers divide the remaining labour as follows: \citet{Gomes_nonequiv} proves the categorical non-equivalence of the two formulations and develops the modal aspect of the charge-quantisation contrast; \citet{Gomes_globalform} treats the global form of the Standard Model gauge group, showing how it gives abductive arguments for unification. The derivations of Sections~\ref{sec:Higgs}--\ref{sec:Yukawa} take only the fundamental bundle data as input and are independent of both companions, whose results enter only the comparative discussion of Section~\ref{sec:defense}. 

Section~\ref{sec:PFB_VB} reviews both formulations. Sections~\ref{sec:Higgs}--\ref{sec:Yukawa} rework the Higgs and Yukawa mechanisms from the geometry-first standpoint. Section~\ref{sec:defense} sets out what the tighter alignment excludes, and what it accommodates only at some explanatory cost. Section~\ref{sec:conclusions} concludes.

\section{Symmetry-first and geometry-first formulations of gauge theory}\label{sec:PFB_VB}

I begin with the familiar symmetry-first formulation, then introduce the geometry-first alternative. A word on notation: $\varpi$ always denotes the principal connection one-form on $P$; $\nabla$ denotes a covariant derivative on a vector bundle $E$; and $\omega$ denotes the local connection form on $E$ that appears upon choosing a frame over $U\subseteq M$, so that $\nabla=\d+\omega$ with $\omega\in\Omega^1(U,\mathsf{End}(E|_U))$. When I speak of the \emph{affine structure} of a vector bundle, I mean the space of covariant derivatives compatible with the fibre structure (e.g.\ Hermitian metric): this space is an affine space modelled on $\Omega^1(\mathsf{End}(E))$, so that any two compatible covariant derivatives differ by an endomorphism-valued one-form.

\subsection{Gauge theory and principal fibre bundles: the symmetry-first formulation}\label{sec:PFB}

I call any formulation of gauge theory that introduces symmetries via principal bundles the \emph{principal bundle point of view} (PFB-POV). The construction is standard (see Appendix~\ref{app:PFB} for conventions): a principal $G$-bundle $(P,M,G)$ with connection $\varpi$ determines, for each representation $\rho:G\to GL(V)$, an associated vector bundle
\begin{equation}\label{eq:AVB}
E:=P\times_\rho V, \qquad (p,v)\sim(p\!\cdot\! g,\,\rho(g^{-1})v),
\end{equation}
together with a covariant derivative $\nabla$ induced by $\varpi$ via \eqref{eq:PFB_cov}. Matter fields are sections of these bundles and representation labels encode internal quantum numbers. The same connection $\varpi$ coordinates parallel transport across all associated bundles, it is what ensures that different matter fields charged under the same interaction march in step (see \citep{Weatherall2016_YMGR} and Figure~\ref{fig:associated-bundles}). It is the primacy of the postulated structure group $G$ that makes this symmetry-first.

\begin{figure}[h]
\centering
\[
\begin{tikzcd}[column sep=large,row sep=large]
 & & E_i:=P \times_{\rho_i} V_i ,\,\,,\nabla_i\\
P(M,G,\varpi) \arrow[urr, "\rho_i"] \arrow[rr, dotted] \arrow[drr, "\rho_j"']
 & & \vdots \\
 & &E_j:= P \times_{\rho_j} V_j,\,\,,\nabla_j
\end{tikzcd}
\]
\caption{A principal $G$-bundle $P(M,G,\varpi)$ and its associated vector bundles $E_i:=P\times_{\rho_i}V_i$, where $\rho_i: G\rightarrow GL(V_i)$ is the representation determined by a particle's quantum numbers. Each covariant derivative $\nabla_i$ is induced by $\varpi$ via \eqref{eq:PFB_cov}. See Appendix~\ref{app:PFB}.}
\label{fig:associated-bundles}
\end{figure}

Note, first, that $\rho$ need not be faithful, and its image need not exhaust $\mathsf{Aut}(V)$: the structure group of $P$ and the automorphism group of the typical fibre need not be isomorphic. This gap between $G$ and $\mathsf{Aut}(V)$ is what I call \emph{slack}, and it will be  analysed in Section~\ref{sec:slack}.

Second, note that though a common principal bundle coordinates the covariant derivative on its associated bundles, it does not define unique isomorphisms between such bundles. \footnote{ The candidate map $[\sigma,v]_1\mapsto[\sigma,v]_2$ (suggested to me by Jim Weatherall) is well-defined only when $\rho_1=\rho_2$. For otherwise, under a change of representative, $[g,v]_1=[\sigma\cdot g,\rho_1(g^{-1})v]_1\mapsto[\sigma\cdot g,\rho_1(g^{-1})v]_2\neq [g,v]_2$, since the equality would require  $\rho_2(g)$ for $E_2$. For $G=U(1)$ with $\rho_i(e^{i\theta})=e^{in_i\theta}\mathbb{1}$, it fails whenever $n_1\neq n_2$.} Sharing a principal bundle therefore supplies no relations among distinct matter bundles; those must come from equivariant maps introduced separately. This will be important for Yukawa interactions, treated in Section~\ref{sec:Yukawa}, since these involve different matter bundles. 

\subsection{Gauge theory and vector bundles: the geometry-first formulation}\label{sec:VB-POV}
Again, the geometric perspective I develop here dispenses with the principal bundle altogether. The analogy with spacetime clarifies the aim. Consider $(M,g,\Xi_i)$: a smooth Lorentzian manifold $(M,g)$ with various tensor fields $\Xi_i$ on $M$, living in spaces constructed from the tangent bundle $TM$. The automorphism group of a typical fibre $(T_xM,g_x)$ is $O(3,1)$ (or $SO(3,1)$ if orientation is background structure). This group becomes explicit once we introduce orthonormal frames. Yet we can say much about $g$ and $\Xi_i$ in a purely geometric, frame-independent manner, without ever mentioning $SO(3,1)$. And if we were to posit a different group acting on $TM$ ($O(2)$, say, rather than $SO(3,1)$), we would need a geometric rationale. 

In gauge theory, by contrast, an analogous “frame-free” formulation for the behaviour of matter remains largely undeveloped (cf. \citep{Weatherall2016_YMGR, Gomes_internal}), and the very idea of a geometric interpretation of the groups and their representations is seldom raised. I am after a formulation of gauge theories for which these interpretations are transparent. 

I call the realisation of the geometry-first approach in the context of particle physics the \emph{vector bundle point of view} (VB-POV).\footnote{Other theories could have geometry-first formulations (those based on Cartan geometry, for instance), but my focus is particle physics, where the relevant spaces are vector bundles.} To see what the view must replace, consider the work the principal bundle does in the standard formulation. The main role of $\varpi$ in $(P, M, G)$ is to coordinate covariant derivatives between associated bundles; yet the principal connection itself does not represent any physical field. As \citet[p.~41]{Jacobs_PFB} argues, only the induced connections on the associated bundles carry physical content; yet the coordination between those connections calls for explanation.\footnote{Jacobs’s formulation: ``Neither the principal bundle nor the [principal] connection on its own represent anything physical. Rather, it is the induced connection on the associated bundle that represents the Yang-Mills field. [But] This approach has difficulties in accounting for distinct matter fields coupled to the same Yang-Mills field.’’ He defends instead an `inflationary approach’ that reifies the bundle of connections; I argue against this in \citep{Gomes_internal}.}

In \citep{Gomes_internal} I showed that the coordination has a simpler source: \textit{principal bundles are unnecessary if interacting particles are all sections of the same vector bundles or of their tensor products}. A covariant derivative on $E$ induces one on every tensor product of $E$ and its dual $E^*$, since for $\kappa\in \Gamma(E)$ and $\xi\in \Gamma(E^*)$:
\be  d(\langle \xi, \kappa \rangle)(X) = \langle \nabla_X^* \xi, \kappa \rangle + \langle \xi, \nabla_X \kappa \rangle,
\ee where angle brackets represent contraction. The generalisation to arbitrary tensor products is straightforward. Parallel transport therefore marches in step across all tensor-product bundles; the tensor structure itself is the `common cause’.

The idea is to postulate a family of independent \emph{fundamental} vector bundles $E_1,\ldots,E_k$, from which all further structure is derived. Every field is a section of a tensor product of these bundles and their duals (e.g.\ $\Gamma(E_1\otimes E_1\otimes E_j\otimes E_k^*)$), and the corresponding covariant derivatives $\nabla_1,\ldots,\nabla_k$ furnish the entire dynamical content of a gauge theory expressible from the VB-POV.\footnote{\citet{BhallaLadd_gluon} pose a dilemma for this formulation: either gauge bosons are sections of vector bundles, in which case representing them requires a background flat derivative operator---naturally interpreted as surplus structure---or they are not. In the second horn, they suggest that the contrast with the principal-bundle formulation erodes. I reject the first horn, for the natural interpretation of the force field is $\nabla$ itself: a point of the affine space of compatible covariant derivatives, determined by the fundamental bundle. No background operator needs to be introduced, though only differences $\nabla-\nabla'$ are sections, i.e.\ elements of $\Omega^1(\mathsf{End}(E))$. The second horn needs first to be refined by distinguishing two senses of `gauge boson'. As the primitive field, the gauge boson is $\nabla$, and is not a section. As the particle of perturbative field theory, it is the excitation $\omega=\nabla-\nabla_0$ about a physically significant reference $\nabla_0$, in the regime where $\omega$ is small---and is a section. The two senses peacefully co-exist: particle-talk picks out from the primitive field a background-relative excitation, exactly as $h_{\mu\nu}=g_{\mu\nu}-\eta_{\mu\nu}$ does in the gravitational case, where the background-relativity of $h_{\mu\nu}$ undermines neither the physical standing of gravitational radiation nor the background-independence of the exact theory. Thus, taking $\nabla$ as primitive does not erode the contrast between the formulations, which never was `sections versus connections': it concerns what fixes the group (Section~\ref{sec:defense}).} The class of bundles reachable from a fundamental bundle is closed under finite direct sums, tensor products, duals, and (anti)symmetrised/exterior powers, with the connection induced functorially in each case.

The resulting picture, for a single fundamental vector bundle, is summarised in Figure \ref{fig:VB-POV}.
\begin{figure}[h]
\centering
\[
\begin{tikzcd}[column sep=large,row sep=large]
 & & E\otimes E\otimes E, \nabla\\
E(M,V,\nabla) \arrow[urr, "\otimes_{(3,0)}"] \arrow[rr, dotted] \arrow[drr, "\otimes_{(i,j)}"'] 
 & & \vdots \\
 & &\underbrace{E\otimes...\otimes E}_i\otimes\underbrace{E^*\otimes...\otimes E^*}_j, \nabla
\end{tikzcd}
\]\caption{An example of the kinds of bundles that can be formed from a fundamental vector bundle. Each such tensor product (which could include arbitrary symmetrisations) inherits an affine structure from $\nabla$, and each corresponds to a representation of $\mathsf{Aut}(V)$. In the VB-POV, particle types must arise through this kind of structure.}\label{fig:VB-POV}
\end{figure}

On this view, no gauge groups need be postulated at the ground level. The automorphism groups of the 
fundamental vector bundles, $\mathsf{Aut}(E_n)\subset \mathsf{End}(E_n)$, are already implied by 
their internal structure, and the overall gauge group is simply the product 
$\prod_n\mathsf{Aut}(E_n)$.\footnote{ See \citep[Ch. 7]{Bleecker} for how to 'splice' together the principal bundles with different structure groups.} Should principal fibre bundles be invoked at all, they are entirely 
\emph{derived from} the structure of the vector bundles, which form the fundamental data. 
The familiar distinction between Abelian and non-Abelian theories then appears as a distinction 
between different types of automorphism groups. In particular, one-dimensional vector bundles, 
whose typical fibres are isomorphic to $\mathbb{C}$, generate Abelian automorphism groups.  

Matters of gauge invariance are now also seen under a different lens. Any composite object constructed from 
the basic dynamical variables (the $\nabla_n$ and other tensor fields) will be tensorial, that is, 
covariant under the corresponding automorphism groups; and any scalar formed from such quantities 
will be invariant under those groups. In this respect the description parallels the spacetime case that opened this subsection: modern abstract-index treatments of general relativity scarcely mention ``gauge invariance'' or ``coordinate invariance''; covariance is secured at the ground level.\footnote{Upon quantisation, as I argue in \citep{Rep_conv}, 
superpositions of states may require relating objects across distinct classical possibilities. 
That, I contend, is where ``gauge fixing''---or, more broadly, what I call 
\emph{representational schemes}---enters. Gauge fixings become necessary when a fixed reference 
across classical states is needed. If such references are physical, they can, incidentally, be used 
to describe content in a gauge-invariant (or gauge-fixed) manner, in the traditional PFB-POV sense.\label{ftnt:QM}}

This vantage point also reframes the question that closed Section~\ref{sec:PFB}: canonical maps between the PFB-POV's associated bundles required $\rho_1=\rho_2$; here, all vector bundles charged under a given interaction ascend from a single \emph{fundamental} bundle. In the cases to be explored below, each such fundamental bundle
$E_n$ has typical fibre $\mathbb{C}^n$ and is equipped with an inner product $\langle\,\cdot\,,\,\cdot\,\rangle$
and, where appropriate, a unit complex volume form~$\epsilon$, a trivialisation of the determinant line, which I will also call a complex orientation.\footnote{With the caveat that the term is an analogy: every complex vector space already carries a canonical \emph{real} orientation; what a volume form $\epsilon$ fixes is the complex determinant, reducing $U(n)$ to $SU(n)$.} The various associated 
bundles then appear not as independently defined objects requiring ad~hoc identifications, but as 
constructions from $E_n$ itself. Their mutual relations are fully accounted for by the 
standard functorial machinery of geometry. For instance, an element of $E^n$ contracts with an element of $E^{n*}\wedge E^{n*}\otimes E^{n}$ via the interior product on the exterior factors, or the inner product between the two copies of $E^n$. The virtue of the geometric route is that it trades purely on geometrical language; the mere availability of a formulation that sidesteps algebraic machinery is already significant.

The Standard Model fits neatly within the geometry-first picture (the scope and limitations of the formulation are assessed in Section~\ref{sec:defense}). In the standard PFB-POV every particle field is a section of an associated bundle for a principal bundle with structure group $SU(3)\times SU(2)\times U(1)$, and the representations that appear are tensor products of fundamental representations. The geometry-first alternative is available because the corresponding associated bundles can be constructed geometrically from fundamental vector bundles---via tensor and exterior products, (anti)symmetrisation, determinants, and the like. Namely, we fix once and for all the three fundamental vector bundles (cf. \cite{Gomes_internal}):
\be\label{eq:fund_bundle_data}
\big(E^3,M, \bb{C}^3, \langle\cdot,\cdot\rangle_3,\epsilon_3\big),\qquad
\big(E^2, M, \bb{C}^2,\langle\cdot,\cdot\rangle_2,\epsilon_2\big),\qquad
\big(E^1, M, \bb{C},\langle\cdot,\cdot\rangle_1\big).
\ee
I will refer to \eqref{eq:fund_bundle_data} as the \emph{fundamental bundle data}. The gauge group in the VB-POV is then determined, fibrewise, by
\be\label{eq:G_VB}
G_{\mathrm{VB}}:=\mathsf{Aut}(E^3_x,\langle\cdot,\cdot\rangle_3,\epsilon_3)\times\mathsf{Aut}(E^2_x,\langle\cdot,\cdot\rangle_2,\epsilon_2)\times\mathsf{Aut}(E^1_x,\langle\cdot,\cdot\rangle_1)\cong SU(3)\times SU(2)\times U(1),
\ee
where each factor is the structure-preserving automorphism group of the corresponding typical fibre. A covariant derivative \textit{on a single} vector bundle suffices to encode \textit{one} fundamental interaction.\footnote{\label{ftnt:gauge_group}There are two levels of automorphism groups: the finite-dimensional group $\mathsf{Aut}(E_x)$ of a single structured fibre, and the group $\mathsf{Aut}(E)$ of vertical structure-preserving bundle automorphisms (equivalently, sections of the associated group bundle), which is the (infinite-dimensional) gauge group $\mathcal{G}(E)$. Statements such as `the gauge group is $SU(3)\times SU(2)\times U(1)$' are fibrewise statements, about $\mathsf{Aut}(E_x)$; wherever $\mathsf{Aut}(E)$ appears without a subscript (as in Proposition~\ref{prop:param}), the vertical automorphism group is meant, acting fibrewise through $\mathsf{Aut}(E_x)$.}

One methodological remark before the derivations. Sections~\ref{sec:Higgs} and~\ref{sec:Yukawa} take only the fundamental bundle data and their compatible covariant derivatives as input; the mass form, the electroweak spectrum, and the Yukawa contractions are computed from these without symmetry postulates. At several points the text takes brief detours to recover symmetry-first structures within the new formulation: unitary gauge (Section~\ref{sec:Higgs_dyn}), the unbroken subgroup (Section~\ref{sec:SU2}), weak isospin and the Higgs frame (Section~\ref{sec:Yukawa2}), and, at length, the comparative apparatus of Section~\ref{sec:defense}. These are mere translations: the derivations do not depend on them, and where they seem contrived, it is because the symmetry-first vocabulary is not native to the objects that feature in the geometry-first descriptions. The dictionary, in brief, is as follows:
\begin{center}
\begin{tabular}{l|l}
\textbf{symmetry-first} & \textbf{geometry-first}\\
\hline
weak doublet / singlet & matter bundle with / without an $E^2$ factor (cf.\ \eqref{eq:assignments})\\
hypercharge $Y$ & tensor power of $E^1$: $n=3Y$, duals counting negatively\\
gauge potential & difference of covariant derivatives, $a=\nabla-\nabla^{(0)}$\\
unbroken Lie algebra & kernel of $X\mapsto X\,e_0$ (Lemma~\ref{lem:kernel})\\
Goldstone directions & orbit directions never coordinatised (Proposition~\ref{prop:param})\\
conjugate Higgs $\tilde\phi=i\sigma_2\phi^*$ & bundle-level conjugate $\boldsymbol{\varphi}_c=\widetilde C(\boldsymbol{\varphi})$ (Section~\ref{sec:Yukawa2})\\
invariant Yukawa coupling & canonical contraction of the declared structures\\
\end{tabular}
\end{center}

I now apply the VB-POV to the Higgs mechanism; the more familiar PFB-POV treatment can be found in any standard textbook (see e.g. \cite[Ch. 10.3]{Bleecker}).

\section{The Higgs mechanism in the geometry-first formulation}\label{sec:Higgs}

%The proof, they say, is in the eating of the pudding. So here, to prove that the geometry-first perspective embodied by the VB-POV is sufficiently different to the PFB-POV to merit attention, I will provide a stand-alone derivation of the (classical) Higgs mechanism. 

In the standard presentation (see, e.g.,~\cite[Ch.~2]{Tong_SM},~\cite[Ch.~10.3]{Bleecker}), the Higgs mechanism is usually described in the language of spontaneous symmetry breaking. One is then compelled to invoke stabilisers, quotients of groups, Killing forms on Lie algebras,  Goldstone’s theorem, among other tools. In this section I outline an equivalent treatment, differing only in that it is phrased entirely in the language of vector bundles; the essential structure becomes visible without appeal to symmetry. It also shows that avoiding explicit symmetry need not produce unwieldy proofs (a common objection to some reformulations; e.g.\ the Einstein algebra formulation of general relativity \citep{Geroch1972}). 

\paragraph{Remark (Structure of the Higgs bundle).}
The fundamental bundle data~\eqref{eq:fund_bundle_data} specifies
$E^2$ with Hermitian metric \emph{and} orientation~$\epsilon_2$, so that
$\mathsf{Aut}(E^2_x)\simeq SU(2)$; and $E^1$ with Hermitian metric alone, so
that $\mathsf{Aut}(E^1_x)\simeq U(1)$.  The Higgs field, however, is a section
not of~$E^2$ but of a tensor product $E^2\otimes (E^1)^{\otimes n}$, where
$n$ encodes the hypercharge (cf.~Table~\ref{tab:fermion-reps}).

This tensor product inherits a Hermitian metric from those on $E^2$
and~$E^1$, but it does \emph{not} inherit an orientation: the determinant
line bundle satisfies
$\Lambda^2\!\big(E^2\otimes (E^1)^{\otimes n}\big)\simeq (E^1)^{\otimes 2n}$,
which is not trivialised by $\epsilon_2$ alone.  Hence the
structure-preserving automorphism group of each fibre is $U(2)$, not $SU(2)$.
Geometrically, the orientation of~$E^2$ and the phase freedom
of~$(E^1)^{\otimes n}$ are entangled through the central
$\mathbb{Z}_2$ in $U(2)\simeq (SU(2)\times U(1))/\mathbb{Z}_2$.
%(Here the $U(1)$ is the effective phase group of $(E^1)^{\otimes n}$; pulled back to the phase group of $E^1$ itself, the kernel is $\mathbb{Z}_{2n}$---for $n=3$, the $\mathbb{Z}_6$ of Section~\ref{sec:composite}.)

This is why the treatment below, which requires only a Hermitian bundle
without orientation, applies directly to the electroweak Higgs.  For the
mechanism itself, the Higgs field sees only the $U(2)$ structure of
$E^2\otimes (E^1)^{\otimes n}$; any two unit sections are related by
a $U(2)$ transformation, and the specific tensor power $n$ plays no role.
The specific tensor power does no work in the mechanism itself: it is fixed ($n=3$ for the electroweak Higgs) only by the Higgs's couplings to other sectors via Yukawa terms (Section~\ref{sec:Yukawa}), where cross-sector charge balance determines it uniquely.  The unit complex volume form $\epsilon_2$ is likewise invisible
to the Higgs mechanism; it enters only through the conjugate-Higgs
construction needed for the quark Yukawa coupling (Section~\ref{sec:Yukawa2}).

\subsection{The  Higgs field as a background}\label{sec:Higgs_field}
Let $(E, M, \mathbb{C}^n, \langle \cdot,  \cdot\rangle)$ be a Hermitian vector bundle over a Lorentzian manifold $(M, g)$, and let $\nabla$ be a covariant derivative on $E$ compatible with the Hermitian structure. (I drop the subscript $n$ when it is clear from context.) Write $ \|\kappa(x)\|^2:=\langle \kappa(x),\kappa(x) \rangle$ for $\kappa \in \Gamma(E)$. As usual, the norm of a tensor product factorises linearly, i.e. for  $\lambda\in  \Gamma(TM)$ and $\xi\in \Gamma(E)$, we have: 
\be \|\lambda \otimes \xi\|^2:=g(\lambda, \lambda)\langle \xi, \xi\rangle.\label{eq:fac_ip}
\ee
Thus, when taking norms of tensors that are not spacetime scalars, $g$ is used for the spacetime (abstract) indices, but I will leave it implicit, absorbing it into $\langle\,.\,,\,.\,\rangle$; being built from a Lorentzian $g$, such `norms' are of course indefinite in their spacetime indices.

As further background structure, fix $\varphi \in \Gamma(E)$ with everywhere non-zero norm, $\|\varphi(x)\| >0$; write $\Gamma_0(E)$ for this nowhere-zero sector of $\Gamma(E)$.\footnote{\label{ftnt:zero_sector}The construction lives on this sector, or locally on the complement of the zero set of $\varphi$: it describes perturbation theory around the electroweak vacuum. Zeros of $\varphi$ can carry non-trivial winding, in which case $e_0=\varphi/\|\varphi\|$ does not extend across them; such defect sectors require a separate, patchwise analysis. A nowhere-vanishing section is also a global reduction of the bundle, so restricting to $\Gamma_0(E)$ is a genuine restriction, not a regularity convention.}  Picking out a constant $v>0$ (later to be the minimum of the potential), we write $\|\varphi(x)\| =(H(x)+v)$, for $H \in C^\infty(M)$, and get
\be
\label{eq:varphi1}\varphi(x) =\|\varphi(x)\| e_0(x)= (H(x) + v) e_0(x),
\ee
where $e_0(x)=\frac{\varphi(x)}{\|\varphi(x)\| }$ is a unit section i.e. $\langle e_0(x), e_0(x) \rangle = 1,\,\,\forall x\in M$ which is  well-defined since  $\|\varphi(x)\| >0,\,\,\forall x\in M$ (the $x$-dependence is suppressed from now on).

The mechanism is already visible in the Higgs kinetic term:
\be {L}_{\text{kin}, \varphi}(\nabla)=\int_M  \mathcal{L}_{\text{kin}, \varphi}=\int_M \d^4 x\, \langle\nabla\varphi, \nabla\varphi\rangle,
\ee
where the calligraphic $\mathcal{L}$ will denote Lagrangian densities. 

Here the Higgs is treated as a background structure. But its status is a
halfway house. The covariant derivative is not required to preserve
$\varphi$: that failure is precisely what \eqref{eq:Higgs_cov} measures.
But neither does $\varphi$ yet have dynamics of its own; the potential, and
with it the Higgs's own mass, enters only in Section~\ref{sec:Higgs_dyn}. At this stage I only need the mechanism for mass acquisition of the vector bosons; for the connection to be dynamical, the Lagrangian must also include the standard Yang--Mills kinetic term ($\mathcal{L}_\nabla=\mathsf{Tr}(\Omega\wedge *\Omega)$, cf.\ \eqref{eq:Lag_Omega}).

Since $e_0$ has unit norm,
\be 0=\label{eq:ortho_e}\nabla\langle e_0, e_0 \rangle=2\mathsf{Re}\, \langle e_0, \nabla e_0\rangle, \quad\text{and}\quad \nabla v=0,
\ee
where $\mathsf{Re}$ takes the real component.\footnote{The induced inner product that appears upon expansion of quadratic terms of the form $\|\xi\|^2$ is $\text{Re}\langle \cdot, \cdot \rangle$, which is effectively what appears in Lagrangians, due to the use of the complex conjugate terms, cf. \cite[Ch. 8]{Hamilton_book}. \label{ftnt:ip}} Using \eqref{eq:varphi1} and  \eqref{eq:ortho_e} we get:
\be\label{eq:Higgs_square}
\langle \nabla \varphi, \nabla \varphi \rangle=\|\nabla \varphi\|^2 = (\partial H)^2 + (H+v)^2 \langle \nabla e_0, \nabla e_0 \rangle,
\ee
where $\partial$ is the exterior derivative  acting on scalars, so $(\partial H)^2=g^{\mu\nu}\partial_\mu H\,\partial_\nu H$.

There are no cross-terms between $\nabla e_0$ and $e_0$. The covariant derivative appears quadratically in $(H +v)^{\,2}\langle \nabla e_0, \nabla e_0 \rangle$; when $H$ is small relative to $v$, such terms endow mass to the affine structure, as I now show.

First, note that $\nabla e_0$ does not contain all the information in $\nabla$:
the components of $\nabla$ that do not enter $\nabla e_0$ are absent from the Higgs kinetic term, and so receive no mass contribution from it: they remain `massless' as far as this Lagrangian term is concerned.\footnote{This does not mean those components of the connection are trivial or flat; they may well have nontrivial curvature and dynamics sourced by other terms (e.g.\ the Yang--Mills term $\mathsf{Tr}(\Omega\wedge *\Omega)$). The point is only that the Higgs kinetic term does not generate a quadratic (mass) term for them.}
From \eqref{eq:ortho_e}, only the components of $\nabla$ that rotate $e_0$ into directions orthogonal to $e_0$
appear in the quadratic term, while those preserving the real vector space generated by $e_0$ drop out. In more detail, from \eqref{eq:ortho_e}, the relevant inner product arising from the squared norm is $\mathsf{Re}\, \langle\,.\,,\,.\, \rangle$ (see footnote \ref{ftnt:ip}). Using it we  project $\nabla e_0$ along $e_0$ and its orthogonal complement to find a geometric description of \eqref{eq:Higgs_square} as follows. 

The unit section \(e_0\) defines both a complex line \(L_{\mathbb{C}}:=\mathrm{span}_{\mathbb{C}}\{e_0\}\subset E\) and a real line \(L_{\mathbb{R}}:=\mathrm{span}_{\mathbb{R}}\{e_0\}\); the two will play different roles below. Because of \eqref{eq:ortho_e}, here we need the orthogonal complement of \(e_0\) with respect to $\text{Re}\langle \cdot, \cdot \rangle$, call it \(L^{\perp}\); it contains $ie_0$, and is interpreted as a normal bundle to \(L_{\mathbb{R}}\) within \(E\). The covariant derivative
\(\nabla e_0\) then defines a one-form with values in \(L^{\perp}\); it probes how \(L^{\perp}\) curves within $E$. In standard sub-bundle language this is the second fundamental form of the line $L_{\mathbb{R}}$, evaluated on its unit section; I will call it a mixed
\emph{shape operator} for \(L^{\perp}\):\footnote{In differential geometry, given a hypersurface $S\subset M$ with unit normal $n$, the standard shape operator for $S$ is defined as a tensorial operator  $K_n:TS\rightarrow TS$, namely $K=\nabla n$. Here  our operator mixes indices, $K_{e_0}:TM\rightarrow L^{\perp}$ (it is the second fundamental form $B_L(X,s):=P^\perp\nabla_X s$ of $L_{\mathbb{R}}\subset E$ at $s=e_0$), but it can still be thought of as describing the shape of $L^{\perp}$ within $E$ as one goes around $M$; hence the name.}
\be
K_{e_0}(X)\;:=\;P^\perp_{e_0}\,\nabla_X e_0\;=\;(\nabla_X e_0)^\perp=\;\nabla_X e_0
\ee
where \(P^\perp_{e_0}:=\mathsf{Id}-\mathsf{Re}\, \langle\,e_0\,,\,.\, \rangle e_0\) is the orthogonal projection onto \(L^{\perp}\) and $X\in TM$.
From   \eqref{eq:ortho_e} we get
\be 
\langle \nabla e_0,\nabla e_0\rangle=\;\langle (\nabla e_0)^\perp,(\nabla e_0)^\perp\rangle\;=\;\langle K_{e_0},K_{e_0}\rangle=\|K_{e_0}\|^2,
\ee
i.e. the norm squared of the second fundamental form, or extrinsic curvature, defined by $e_0$.

The operator $\nabla$ merely provides the affine structure of the bundle, while the 
Higgs field singles out an internal direction $e_0$. From the kinetic term, assuming $H$ small compared to $v$, we obtain:
\be\label{eq:Higgs_cov}  \langle \nabla \varphi, \nabla \varphi \rangle\approx  v^2\|K_{e_0}\|^2.
\ee
That is, the Higgs kinetic term induces a positive semidefinite quadratic form on the space of compatible covariant derivatives: the affine structure `acquires mass' precisely in the directions orthogonal to the Higgs; the mass term is the squared norm of the extrinsic curvature of the corresponding orthogonal sub-bundle, weighted by the squared vacuum expectation value $v^2$. Note that Equation~\eqref{eq:Higgs_cov} is expressed entirely in abstract-index (tensorial) language, with no appeal to symmetries or their breaking. I am, of course, `breaking the symmetry' by postulating a background field $\varphi$; but any two non-zero Higgs fields define structurally equivalent directions (Section~\ref{sec:Higgs_dyn}), so only the non-zero norm $v$ matters.

Two points of contact with the standard formalism. First, `mass acquisition' is, at this stage, a statement about the quadratic form \eqref{eq:Higgs_cov}, not about gauge bosons in the particle-physics sense; identifying specific massive and massless components requires writing $\nabla=\d+\omega$ in a frame adapted to $e_0$ (Section~\ref{sec:SU2}). Second, the decomposition presupposes $\varphi$ nowhere vanishing---the sector restriction of footnote~\ref{ftnt:zero_sector}.

The geometric picture also makes vivid a fact that the symmetry-first formulation tends to obscure. After symmetry breaking, the Higgs vacuum singles out the complex line $L_{\mathbb{C}}=\mathrm{span}_{\mathbb{C}}\{e_0\}\subset E$, splitting $E\cong L_{\mathbb{C}}\oplus L_{\mathbb{C}}^{\perp}$ (the complex orthogonal complement). But there is no reason for the covariant derivative to preserve this splitting. That is, let $P_{L_{\mathbb{C}}}$ be the orthogonal projector onto $L_{\mathbb{C}}$, then,  in general, $\nabla P_{L_{\mathbb{C}}}\neq 0$. 
Alternatively, in a local frame adapted to the splitting, the connection one-form need not be block-diagonal, and the off-diagonal components measure the failure of $L_{\mathbb{C}}$ and $L_{\mathbb{C}}^{\perp}$ to be parallel subbundles. Hence parallel transport around a closed loop can mix the two components. Applied to the electroweak theory (Section~\ref{sec:Yukawa}), this means that a left-handed electron, defined as the component of $\mathbf{L}_L$ along the Higgs direction, can return from a loop as a mixture of electron and neutrino. While this is almost certainly known to experts,  it is not something one reads off from the unbroken $U(1)_{\mathrm{EM}}$ alone; for in the symmetry-first presentation, the unbroken subgroup preserves the splitting \emph{by definition}, so the mixing must be rescued from the broken generators. The geometry-first picture makes this possibility completely transparent: \emph{the splitting is a condition on the section, not on the connection}.\footnote{This remains true even on topologically non-trivial backgrounds: topology governs whether the splitting $E\cong L_{\mathbb{C}}\oplus L_{\mathbb{C}}^{\perp}$ exists globally, whereas its preservation is a further differential condition on the connection.}

The point bears on a dictum of Coleman's, recently recalled by \citet{Susskind2026}: he used to say  the Higgs effect `leaves no track' (no Goldstone bosons, no degeneracy in the spectrum), so a broken gauge symmetry would be indistinguishable from no symmetry at all. The dictum is true of the spectrum and false of the theory's affine structure. Parallel transport around a closed path  can return matter transformed by the  symmetry that is concealed by the spectrum.%; the track sits in $K_{e_0}=\nabla e_0$: its component normal to the complex Higgs line produces the charged-current mixing, and its $ie_0$-component records the massive neutral phase direction.
\footnote{Configurations with such transformed returns are string-like. They are not topologically protected in the electroweak theory, whose vacuum manifold is a three-sphere: holonomy in the broken directions is physical.} This shows the term `broken' misleads twice over. If we read `broken' as the claim that the gauge structure below the electroweak scale is the reduced pair of unbroken subgroup and compatible connection, it erases `the track': a $U(1)_{\mathrm{EM}}$ connection has only electromagnetic holonomy, which cannot mix electron and neutrino. If we read `broken' as the claim that the splitting $E\cong L_{\mathbb{C}}\oplus L_{\mathbb{C}}^{\perp}$ is a fixed background structure,  compatible connections would satisfy $\nabla P_{L_{\mathbb{C}}}=0$ and transport would be block-diagonal; in this case there is no `track' to begin with.  This is why Coleman's own name for the phenomenon, `secret symmetry' \citep{Coleman_Aspects}, was more accurate: the symmetry is hidden by a dynamical section, not broken by background structure. %the classical counterpart of the point Elitzur's theorem makes in the quantum theory (Section~\ref{sec:Higgs_dyn}).

\subsection{Dynamical mass generation for the Higgs in the linearised theory}\label{sec:Higgs_dyn}

A `halfway house' background Higgs field sufficed for the qualitative features of mass acquisition, as \eqref{eq:Higgs_cov} shows. I now include the dynamics and make the mechanism more precise in the linearised regime, which allows us to derive the Higgs's own mass.

The background structure is now $(E, M, \mathbb{C}^n, \langle \cdot,  \cdot\rangle)$, without a preferred $\varphi$. The task is to describe the space of physically distinct (non-isomorphic) configurations of the Higgs field: the different physical possibilities for a single section in $\Gamma(E)$.

Assume $E$ admits a unit section and pick one, $e_0(x)$. Any two choices $e_0, e'_0$ are isomorphic: they are structurally indistinguishable within $E$, having the same unit norm and no further background structure to tell them apart.  Of course, were we considering the configuration space for two fields, i.e. $\Gamma(E\oplus E)$, sections $\varphi$ and $\varphi'$ need not be collinear. But for the configuration space of a single field, $\Gamma(E)$, or tensor products thereof, there is no canonical comparison between  two non-isomorphic configurations (apart from their norms). So, in order to parametrise the space of non-isomorphic unit sections, I fix the same $e_0$ across configurations.  This idea is described in full in \citep{Rep_conv, Kabel2025, GomesButterfield_hole2}: they would say $e_0$ defines a `representational scheme' across physical possibilities, so that each configuration $\varphi\in \Gamma(E)$ is parallel to $e_0$.  

Equivalently, in symmetry-first language, any two such choices are related by an element of $\mathsf{Aut}(E)$.  So this step is related to the choice of `unitary gauge' in the standard presentation, and it is the only place where symmetry could be mentioned.\footnote{ This argument requires $\mathsf{Aut}(E_x)\simeq U(n)$, i.e.\ that $E$ carry no orientation.  For the electroweak Higgs, which is a section of $E^2\otimes (E^1)^{\otimes n}$, this is indeed the case: the tensor product inherits only a Hermitian metric, not an orientation, as explained in the Remark at the start of this section. The relationship between representational schemes and gauge-fixings is discussed in \cite[Sec. 3.3]{Rep_conv}.} What survives of `gauge choice' is $e_0$ itself, defined physically by the direction of the vacuum.

Thus the configuration space of non-isomorphic single field configurations  can be parametrised by positive scalar functions as follows:
\be\label{eq:config_space} \Gamma_P(E):=\{\varphi=(H+v)e_0,\,\, H\in C^\infty(M),\,\, H+v>0 \text{ pointwise}\}.
\ee
Here $v$ could be absorbed by $H$, but for what follows it is useful to keep them separate.\footnote{The positivity clause is not a physical constraint: it preserves the reading of $H+v$ as the norm $\|\varphi\|$, as in \eqref{eq:varphi1}, and it holds automatically in the linearised regime $|H|\ll v$. It is, however, what makes the slice non-redundant ($re_0$ and $-re_0$ share their pointwise norm and are related by the automorphism $-\mathbb{1}$), and what keeps it inside the nowhere-zero sector.}

$\Gamma_P(E)$ is an open convex subset of an affine space over $C^\infty(M)$ (open because of the pointwise constraint $H>-v$), so its tangent space at $w_o=ve_0$ is still generated by curves of the form $\gamma(t)=(v+th)e_0$ for fixed $h\in C^\infty(M)$, with $\gamma(0)=w_o$ and $\dot\gamma(0)=h\,e_0$. Hence
\be\label{eq:lin_H} T_{w_o}\Gamma_P(E)=\{ H e_0,\,\, H\in C^\infty(M)\}.
\ee

$H$ parametrises the tangent directions to the space of physically distinct Higgs configurations. Rather than working with tangent bundles to configuration spaces, it is more practical to pass to the linearised regime: $\mathcal{O}(H) = \mathcal{O}(a)= \mathcal{O}(K_{e_0}) = \varepsilon$.\footnote{In the comparative sense: that $\frac{|H|}{v}\sim \varepsilon\ll 1$, and \emph{mutatis mutandis} for $a$. Since covariant derivatives form an affine space, smallness of $\nabla$ needs an origin: choose a compatible background derivative $\nabla^{(0)}$ with $\nabla^{(0)}e_0=0$ (declare $e_0$ parallel on $L_{\mathbb{C}}$, pick any compatible derivative on the complex orthogonal complement $L_{\mathbb{C}}^\perp$, and take the direct sum; the complex splitting is global on the nowhere-zero sector), and write $\nabla=\nabla^{(0)}+a$ with $a\in\Omega^1(M,\mathfrak{u}(E))$ tensorial. Then $K_{e_0}=a\,e_0$, and $\mathcal{O}(a)=\varepsilon$ is well defined. As to the choice of $\nabla^{(0)}$,  any two compatible origins with $\nabla^{(0)}e_0=0$ differ by a one-form annihilating $e_0$, and so define the same Higgs-induced mass form.}
  Inserting $\varphi=(H+v)e_0$ into \eqref{eq:Higgs_square} and truncating at quadratic order:
\be\label{eq:vbmass}
\|\nabla \varphi\|^2 = (\partial H)^2 + v^2\|K_{e_0}\|^2 + \mathcal{O}(\varepsilon^3).
\ee
% (I will omit the description of the space of `infinitesimal' connections, $\omega$, since it is similar.)

The kinetic term for the Higgs is $(\partial H)^2$; $v^2\|K_{e_0}\|^2$ is the mass term for part of the affine structure. For the Higgs's own mass we need the potential.

With the potential included, the full Lagrangian reads (in a kinetic-minus-potential sign convention, which is all the mass computation needs)
\be L(\nabla,\varphi)=\int_M\d^4 x\left( \langle\nabla\varphi, \nabla\varphi\rangle-V(\|\varphi\|)\right)+\int_M \mathsf{Tr}(\Omega\wedge *\Omega).
\ee
The minimum of the potential satisfies
\be \delta V (w_0)=0,
\ee
where $V:\Gamma(E)\rightarrow \bb{R}$ is assumed homogeneous on $M$ and dependent only on $\|\varphi\|$, so that $w_o(x)=v\,e_0(x)\in \Gamma_0(E)$ with $\|w_o(x)\|=v\neq0$. Any potential with a non-zero minimum suffices; the familiar `Mexican hat' $V=-\mu\|\varphi\|^2+\lambda\|\varphi\|^4$ is one such choice. (By a mild abuse, $V$ denotes both the pointwise function of $\|\varphi\|$ and the functional it induces.)

Now let us translate back to the symmetry-first approach. That the parametrisation \eqref{eq:config_space} of $\Gamma_P(E)$  is both exhaustive and non-redundant can be formally translated to the symmetry-first language by  the following proposition; the proof is in Appendix~\ref{app:proofs}.\footnote{The proposition quantifies over the full group of unitary vertical automorphisms of $E$ (footnote~\ref{ftnt:gauge_group}). In the Standard Model application the automorphisms are induced from the fundamental bundles: fibrewise these fill out $U(2)$, and the global lift through $SU(2)\times U(1)\to U(2)$ exists on the topologically trivial sector around the electroweak vacuum that is relevant to perturbation theory. Global-form subtleties are treated in \citet{Gomes_globalform}.}

\begin{prop}[Exhaustiveness and non-redundancy of $\Gamma_P(E)$]\label{prop:param}
Let $(E,\langle\cdot,\cdot\rangle)$ be a Hermitian vector bundle of rank~2 admitting a nowhere-vanishing section, with $\mathsf{Aut}(E_x)\simeq U(2)$ (no orientation fixed),\footnote{Rank 2 is the electroweak case. For rank $n$ over a four-dimensional base the argument is unchanged: $\ell_1^\perp$ and $\ell_0^\perp$ in the proof have equal total Chern class, and complex bundles of rank $\geq2$ with equal Chern classes over a base of dimension $\leq4$ are isomorphic.} and let $e_0$ be a unit section. Then:
\begin{enumerate}[(i)]
\item for every $\varphi\in\Gamma(E)$ with $\|\varphi\|>0$ everywhere, there is a $u\in\mathsf{Aut}(E)$ with $u\varphi=\|\varphi\|\,e_0$;
\item two nowhere-vanishing sections are related by an automorphism if and only if their norms agree pointwise;
\item hence $\Gamma_P(E)$ contains exactly one representative of each $\mathsf{Aut}(E)$-orbit of nowhere-vanishing sections, and the subgroup of $\mathsf{Aut}(E)$ fixing $\Gamma_P(E)$ pointwise is the stabiliser $\mathsf{Aut}(E;e_0)$, acting fibrewise as the $U(1)$ of the orthogonal line $\ell^\perp:=\mathrm{span}_{\mathbb{C}}\{e_0\}^\perp$.
\end{enumerate}
\end{prop}

The end-point is the same parametrisation $\varphi=(H+v)e_0$. Extensionally, $\Gamma_P(E)$ coincides with the image of a unitary-gauge slice.

In comparison, the standard textbook approach leads to an equivocation that I now describe, since it illustrates an explanatory benefit of the geometry-first picture, which avoids it.

 In the standard textbook language, unitary gauge takes the full space $\Gamma(E)$ as the configuration space. Symmetry-breaking then fixes a physical vacuum $w_o$, and one expands $\varphi$ around $w_o$ in all four real directions.  In this presentation the redundant modes are present at an intermediate stage and must be argued away, which one does by identifying three of the real directions as Goldstone modes tangent to the gauge orbit, which then can be removed by a field-dependent gauge transformation, which is the transformation required to achieve unitary gauge. So unitary gauge eliminates the modes that are orthogonal to the Higgs vacuum $w_o$ (the Goldstone modes). But here is the source of equivocation:  inner products are invariant under gauge transformations: they are part of the structure that is preserved. So how can modes characterised by being orthogonal to the vacuum---a gauge-invariant property---be eliminated by a gauge transformation? The implied trick is to transform the Higgs, $\Phi\mapsto g\Phi$, while holding $w_o$ fixed; again, were it not held fixed, $\langle g\Phi, w_o\rangle$ would be left invariant. The gauge transformation changes $\langle g\Phi, w_o\rangle$ precisely because $w_o$ is \emph{not} being co-transformed: it is a gauge-fixing reference, not a physical object. But earlier in the derivation $w_o$ played the role of a physical vacuum whose stabiliser determined the unbroken gauge directions. The standard treatment thus equivocates on the status of $w_o$: first a physical vacuum, then a fixed convention that is not gauge-covariant. In the geometry-first picture of this theory the equivocation does not arise.\footnote{
In truth, only the norm $\|\Phi\|=v$ is gauge-invariant (Section~\ref{sec:Higgs_field}); the quantum theory makes this sharper: by Elitzur's theorem \citep{Elitzur1975}, local gauge symmetry cannot be spontaneously broken, so $\langle \Phi\rangle=0$ in any gauge-invariant state; the physical content of `symmetry breaking' resides entirely in gauge-invariant quantities such as $\langle \|\Phi\|^2\rangle\neq 0$. The norm itself breaks does not break the local symmetry: it is preserved by the automorphisms since these act fibrewise unitarily, and by Proposition~\ref{prop:param}(ii) sections with different norms are not automorphism-related in the first place. A non-zero $v$ selects among physically inequivalent possibilities, not among gauge-related representatives of one possibility.}

In order to get the mass of the Higgs we must similarly expand $V(\varphi)$ up to $\order\varepsilon^2$, thus involving the second variation of $V$ around its minimum,  called \emph{the Hessian}. The Hessian is a symmetric linear map that acts fibre-wise; using the notation $\delta\varphi$ for elements of $T_{w_o}\Gamma_0(E)$ we write it as:
\be \mathsf{Hess}(V)_{w_o}:=\frac{\delta^2 V}{\delta \varphi^1\delta\varphi^2}_{|\varphi=w_o}:T_{w_o}\Gamma(E)\rightarrow T_{w_o}\Gamma(E),
\ee
  where we are taking the minimum to be at $w_o$, with $\delta \varphi^1, \delta\varphi^2\in T_{w_o}\Gamma_0(E)$. 
Thus we have:
\be\label{eq:V_exp} V(w_o+\delta\varphi)=V(w_o)+\frac{\delta V}{\delta \varphi}(w_o)\,\delta\varphi+ \frac12\langle\delta\varphi_1,\frac{\delta^2 V}{\delta \varphi^1\delta\varphi^2}(w_o)\, \delta\varphi_2\rangle+\ldots\,.
\ee

The mass terms will be obtained from \eqref{eq:V_exp} by diagonalising $ \mathsf{Hess}(V)_{w_o}$.  Thus, as in the standard account, the mass terms for different components of the Higgs along the different directions within $T_{w_0}\Gamma_0(E)$  are given by the eigenvalues of the Hessian with those directions as eigenvectors.  But,  per \eqref{eq:lin_H}, in the current case we are considering only $\delta\varphi=H(x) e_0\in T_{w_o}\Gamma_0(E)$: the reduced slice contains only the radial direction. (On the full sector, the Hessian of a potential depending only on $\|\varphi\|$ has vanishing eigenvalues along the orbit directions, the Goldstone counterpart.)  The value of the surviving eigenvalue along $e_0$  will depend on the form of $V(\|\varphi\|)$, which is assumed to depend only on the local values of $\varphi$ and the coupling constants. Here  we will just call this eigenvalue  $m_H^2$.

Since $\frac{\delta V}{\delta \varphi}(w_o)=0$,  using this simplification of the Hessian, we obtain:\footnote{In the standard case where $V=-\mu \|\varphi\|^2+\lambda  \|\varphi\|^4$, using \eqref{eq:config_space} and \eqref{eq:lin_H} we can replace $\|\varphi\|^2\rightarrow (H+v)^2$, and $\frac{\delta}{\delta \varphi}\rightarrow \frac{\partial}{\partial H}$. So $\mathsf{Hess}(V)\rightarrow \frac{\partial^2V}{\partial H^2}$. Then an easy computation reveals: 
\be\label{eq:dv} \frac{\partial V}{\partial H}=-2\mu (H+v)+4\lambda (H+v)^3\ee and
\be\label{eq:d2v} \frac{\partial^2 V}{\partial H^2}=-2\mu +12\lambda (H+v)^2, 
\ee
Setting $ \frac{\partial V}{\partial H}=0$ and $H=0$ in \eqref{eq:dv} we obtain $ v=\sqrt{\frac{\mu}{2\lambda}}$. Setting $v=\sqrt{\frac{\mu}{2\lambda}}$ and $H=0$ in   \eqref{eq:d2v} we obtain $\frac{\partial^2V}{\partial H^2}(w_o)=4\mu=m_H^2.$ The canonically normalised fluctuation $h=\sqrt2\,H$ (Section~\ref{sec:SU2}) then has $m_h^2=\tfrac12 m_H^2=2\mu=4\lambda v^2=2\lambda v_{\mathrm{EW}}^2$, the standard value.
 }
\be \langle\delta\varphi_1,\frac{\delta^2 V}{\delta \varphi^1\delta\varphi^2} \delta\varphi_2\rangle=\langle H e_0, m_H^2 H e_0\rangle=m^2_H H^2,
\ee
 and from  \eqref{eq:V_exp}, up to quadratic order in $\varepsilon$: 
 \be\label{eq:V_exp2} V(\varphi)_{|\varphi=w_o}=V(w_o)+\frac12m_H^2 H^2+\order\varepsilon^3.
\ee

No explicit mention of the apparatus catalogued at the opening of Section~\ref{sec:Higgs} was required (see \cite[Ch.~8.1]{Hamilton_book}; \cite[Ch.~2.2]{Tong_SM}; \cite[Ch.~10.3]{Bleecker} for comparison).

\subsection{The electroweak mass matrix from geometry}\label{sec:SU2}

The mass-acquisition formula~\eqref{eq:vbmass} is abstract, so in this subsection we will extract the standard electroweak spectrum from it. For that, we must describe the real vector bundle on which the shape operator acts.

\paragraph{The real bundle carrying the shape operator.}
Let $E_H:=E^2\otimes (E^1)^{\otimes 3}$ be the electroweak Higgs bundle, with hypercharge $Y=1$ in the units of Table~\ref{tab:fermion-reps}, where $E^1$ carries $Y=\tfrac13$.\footnote{For a section of $E^2\otimes(E^1)^{\otimes n}$ with general $n$, replace $g'B$ by $\tfrac{n}{3}g'B$ throughout; nothing else changes.} The projector $P^\perp_{e_0}$ was defined with respect to $\mathsf{Re}\,\langle\cdot,\cdot\rangle$ (footnote~\ref{ftnt:ip}). Write $E_{\mathbb{R}}$ for the underlying real bundle of $E_H$ (real rank~4), equipped with the Riemannian metric $\mathsf{Re}\,\langle\cdot,\cdot\rangle$. A unit section, locally factorised as $e_0=e_0^2\otimes e_0^1\in E^2\otimes (E^1)^{\otimes 3}$ (nothing below depends on the factorisation being global), determines both a complex line $L_{\mathbb{C}}=\mathrm{span}_{\mathbb{C}}\{e_0\}\subset E_H$ and a real line $L_{\mathbb{R}}=\mathrm{span}_{\mathbb{R}}\{e_0\}\subset E_{\mathbb{R}}$. The section $ie_0$ lies in $L_{\mathbb{C}}$ but is real-orthogonal to $e_0$, since $\mathsf{Re}\,\langle e_0, ie_0\rangle=0$. The shape operator $K_{e_0}=P^\perp_{e_0}\nabla e_0$ therefore takes values in the real rank-\emph{three} bundle
\be\label{eq:L_perp_R}
L^\perp_{\mathbb{R}}\;=\;\big(\ell^\perp\otimes (E^1)^{\otimes 3}\big)_{\mathbb{R}}\;\oplus\;\mathrm{span}_{\mathbb{R}}\{i\,e_0\},
\ee
where $\ell^\perp\subset E^2$ is the complex-orthogonal line to $e_0^2$. The first summand has real rank two; the second is the phase direction of $L_{\mathbb{C}}$. The electroweak mass spectrum is the spectrum of the quadratic form $a\mapsto v^2\|a\,e_0\|^2$ on the four connection directions; its image $a\,e_0$ spans this rank-three bundle. It is computed as follows.

\paragraph{The mass matrix.}
To compare with the textbook spectrum, restore the coupling constants: the compatible covariant derivative on $E^2$ is $\nabla^2=\d-\tfrac{i g}{2}W^a\sigma_a$ in the weak-isospin eigenbasis,\footnote{This uses the standard conventions, not the spin-coefficient basis that would include $e_0$ as a frame element.\label{ftnt:T3}} with $\sigma_a$ the Pauli matrices, and on $E^1$ it is $\nabla^1=\d-\tfrac{ig'}{6}B$, so that the third tensor power carries $-\tfrac{ig'}{2}B$ by the Leibniz rule. With $e_0^2=(0,1)^T$ and $e_0^1=1$:
\be\label{eq:nabla_e0}
\nabla e_0=\Big(-\tfrac{ig}{2}\,W^a\sigma_a-\tfrac{ig'}{2}\,B\,\mathbb{1}\Big)e_0^2
=\begin{pmatrix} -\tfrac{ig}{2}\,(W_1-iW_2)\\[2pt] \tfrac{i}{2}\,(g W_3-g' B)\end{pmatrix},
\ee
so the quadratic term in \eqref{eq:vbmass} is
\be\label{eq:ew_mass}
v^2\|\nabla e_0\|^2=\frac{g^2v^2}{4}\big(W_1^2+W_2^2\big)+\frac{v^2}{4}\big(gW_3-g'B\big)^2.
\ee
On the four real gauge-field directions $(W_1,W_2,W_3,B)$ this is the quadratic form
\be\label{eq:mass_matrix}
\mathcal{Q}=\frac{v^2}{4}\begin{pmatrix}
g^2 & 0 & 0 & 0\\
0 & g^2 & 0 & 0\\
0 & 0 & g^2 & -gg'\\
0 & 0 & -gg' & g'^2
\end{pmatrix}.
\ee
$\mathcal{Q}$ is the coefficient matrix of the mass Lagrangian $\mathbf{A}^T\mathcal{Q}\mathbf{A}$, $\mathbf{A}=(W_1,W_2,W_3,B)^T$; it is not yet the physical mass-squared matrix, for which the real-field convention is $\mathcal{L}_{\mathrm{mass}}=\tfrac12\mathbf{A}^T M^2\mathbf{A}$. Here $v$ is the vacuum value of $\|\varphi\|$, whereas the textbook normalisation writes $\Phi_0=(0,\,v_{\mathrm{EW}}/\sqrt2)^T$; so $v=v_{\mathrm{EW}}/\sqrt2$, the canonically normalised Higgs fluctuation is $h=\sqrt2\,H$, and
\be\label{eq:mass_matrix_phys}
M^2=2\,\mathcal{Q}%=\frac{v_{\mathrm{EW}}^2}{4}\begin{pmatrix}g^2 & 0 & 0 & 0\\0 & g^2 & 0 & 0\\0 & 0 & g^2 & -gg'\\0 & 0 & -gg' & g'^2\end{pmatrix},
\ee
with eigenvalues $(m_W^2,\,m_W^2,\,m_Z^2,\,0)$. Equivalently, reading masses from \eqref{eq:ew_mass} with the standard field normalisations ($m_W^2\,W^+W^-$ for the complex pair $W^\pm=(W_1\mp iW_2)/\sqrt2$, and $\tfrac12 m_Z^2 Z^2$ for the real neutral field) gives, in the charged sector,
\be
m_W=\frac{g v}{\sqrt2}=\tfrac12\,g\,v_{\mathrm{EW}}.
\ee
In the neutral sector the $2\times2$ block of \eqref{eq:mass_matrix_phys} has eigenvalues $\tfrac{v_{\mathrm{EW}}^2}{4}(g^2+g'^2)$ and $0$, with eigenvectors
\be\label{eq:weinberg}
Z=\cos\theta_W\, W_3-\sin\theta_W\, B,\qquad
A=\sin\theta_W\, W_3+\cos\theta_W\, B,\qquad
\tan\theta_W=\frac{g'}{g},
\ee
so that
\be
m_Z=\tfrac12\sqrt{g^2+g'^2}\;v_{\mathrm{EW}},\qquad m_\gamma=0,\qquad m_W=m_Z\cos\theta_W.
\ee
(The numerical values presuppose canonically normalised gauge kinetic terms, which is where $g$ and $g'$ are fixed; the geometry determines the form and its kernel, not the coupling.)

The photon direction generalises. The Higgs kinetic term induces, pointwise on the anti-Hermitian endomorphisms $\mathfrak{u}(E_H)$, the quadratic form (cf.~\eqref{eq:vbmass})
\be\label{eq:mass_form}
q_{e_0}(X,Y)=2v^2\,\mathsf{Re}\,\langle X\,e_0,\,Y\,e_0\rangle,\qquad X,Y\in\mathfrak{u}(E_H);
\ee
a perturbation $a\in\Omega^1(M,\mathfrak{u}(E_H))$ of the covariant derivative enters the mass Lagrangian through $\tfrac12\,g^{\mu\nu}q_{e_0}(a_\mu,a_\nu)$.
\begin{lem}\label{lem:kernel}
$q_{e_0}(X,X)=2v^2\|X\,e_0\|^2\geq0$ (the fibre metric is positive-definite), so the kernel of $q_{e_0}$ is exactly $\mathfrak{g}_{e_0}:=\{X\in\mathfrak{u}(E_H): X e_0=0\}$, the directions annihilating $e_0$ (in symmetry-first terms, the fibrewise infinitesimal stabiliser of the Higgs direction).
\end{lem}
%\noindent (Positivity is a statement about the internal form; contracted with spacetime one-forms through the Lorentzian metric it yields the usual indefinite mass Lagrangian.) Again, it is easy to connect to the symmetry-first formulation: for the electroweak bundle $\mathfrak{g}_{e_0}$ is one-dimensional, generated by the electric charge $Q=T_3+\tfrac{Y}{2}$; the corresponding massless field direction is the combination $A\propto g'W_3+gB$ of \eqref{eq:weinberg}. The massless sector is the stabiliser of the Higgs direction, obtained here as the kernel of a quadratic form rather than through a prior stabiliser analysis.

\paragraph{Back to the geometric reading.}
Translated back into the bundle language, we first decompose
\(\nabla e_0\) according to \eqref{eq:L_perp_R}. Its upper component then takes
values in
\(\ell^\perp\otimes(E^1)^{\otimes 3}\); the corresponding charged weak directions
\(W_1\) and \(W_2\), equivalently \(W^\pm\), rotate the weak component
\(e_0^2\) into its complex-orthogonal complement and thereby account for
the real rank-two summand. The lower component of \(\nabla e_0\) is then proportional to
\(i e_0\): the corresponding neutral massive direction changes only the phase of \(e_0\),
and is therefore complex-collinear with \(e_0\) but orthogonal to it with
respect to the  real inner product.

In this description, both the diagonal weak-isospin connection \(W_3\) and the hypercharge
connection \(B\) act on \(e_0\) through this same real phase direction
\(\operatorname{span}_{\mathbb{R}}\{i e_0\}\). Their non-cancelling
combination \(gW_3-g'B\) moves \(e_0\) along \(i e_0\) and therefore
contributes to the mass quadratic form; after normalisation, this is the
\(Z\)-boson direction. By contrast, the combination \(g'W_3+gB\)
annihilates \(e_0\), so it lies in the kernel identified in
Lemma~\ref{lem:kernel} and remains massless; this is the photon direction.
The Weinberg angle \eqref{eq:weinberg} is the coupling-dependent rotation
of the \((W_3,B)\)-plane that separates this massive direction \(Z\) from
the massless kernel \(A\).

\section{The Yukawa mechanism}\label{sec:Yukawa}

The Higgs mechanism gives mass to gauge potentials; the \emph{Yukawa mechanism} gives mass to matter fields. I follow \cite[Ch.~8]{Hamilton_book}, whose notation is already geometric. Matter fields are sections of $S_L\otimes V$ or $S_R\otimes V$, with $S_{L,R}$ the Weyl spinor bundles and $V$ the internal (representation or tensor-product) factor.  Throughout this section the Lorentz spinor pairing is left implicit, so that the formulas track only the internal contractions; moreover, every Yukawa term is accompanied by its Hermitian conjugate.

\subsection{The standard presentation}\label{sec:Yukawa1}
Left- and right-handed chiral fermions transform in inequivalent representations of $G=SU(3)\times SU(2)\times U(1)$: for instance, $L_L \in (\mathbf{1},\mathbf{2},-1)$ while $e_R \in (\mathbf{1},\mathbf{1},-2)$ (see Table~\ref{tab:fermion-reps}, reproduced from \cite[Table~8.2]{Hamilton_book}). Gauge invariance therefore forbids a bare Dirac mass term $\mathcal{L}_{\mathrm{mass}} = -m\,\bar\psi_L\psi_R$.
\begin{table}[h]
\centering
\begin{tabular}{C{2.2cm} C{3.4cm} C{2.1cm} C{1.9cm} C{1.2cm} C{1.2cm} C{1.2cm}}
\toprule
\textbf{Sector} & \textbf{$SU(2)_L\times U(1)_Y$ rep.} & \textbf{Basis vectors} & \textbf{Particle} & $T_3$ & $Y$ & $Q$ \\
\midrule

\multirow{2}{*}{$Q_L$}
  & \multirow{2}{*}{$\mathbb{C}^2\!\otimes\!\mathbb{C}_{1/3}$}
  & $\begin{psmallmatrix}1\\[1pt]0\end{psmallmatrix}$ & $\mathbf{u}_L$ & $+\tfrac12$ & $+\tfrac13$ & $+\tfrac23$ \\
  & & $\begin{psmallmatrix}0\\[1pt]1\end{psmallmatrix}$ & $\mathbf{d}_L$ & $-\tfrac12$ & $+\tfrac13$ & $-\tfrac13$ \\
\midrule

$u_R$ & $\mathbb{C}\!\otimes\!\mathbb{C}_{4/3}$~~ & $1$ & $\mathbf{u}_R$ & $0$ & $+\tfrac{4}{3}$ & $+\tfrac23$ \\
$d_R$ & ~$\mathbb{C}\!\otimes\!\mathbb{C}_{-2/3}$ & $1$ & $\mathbf{d}_R$ & $0$ & $-\tfrac{2}{3}$ & $-\tfrac13$ \\
\midrule

\multirow{2}{*}{$L_L$}
  & \multirow{2}{*}{$\mathbb{C}^2\!\otimes\!\mathbb{C}_{-1}$}
  & $\begin{psmallmatrix}1\\[1pt]0\end{psmallmatrix}$ & $\nu_{eL}$ & $+\tfrac12$ & $-1$ & $0$ \\
  & & $\begin{psmallmatrix}0\\[1pt]1\end{psmallmatrix}$ & $e_L$ & $-\tfrac12$ & $-1$ & $-1$ \\
\midrule

$e_R$ & $\mathbb{C}\!\otimes\!\mathbb{C}_{-2}$ & $1$ & $e_R$ & $0$ & $-2$ & $-1$ \\
\midrule

\multirow{2}{*}{Higgs $\varphi$}
  & \multirow{2}{*}{~~$\mathbb{C}^2\!\otimes\!\mathbb{C}_{1}$~~}
  & $\begin{psmallmatrix}1\\[1pt]0\end{psmallmatrix}$ & $\varphi^+$ & $+\tfrac12$ & $+1$ & $+1$ \\
  & & $\begin{psmallmatrix}0\\[1pt]1\end{psmallmatrix}$ & $\varphi^0$ & $-\tfrac12$ & $+1$ & $0$ \\
\midrule

\multirow{2}{*}{conj.\ Higgs $\varphi_c$}
  & \multirow{2}{*}{~~$\mathbb{C}^2\!\otimes\!\mathbb{C}_{-1}$}
  & $\begin{psmallmatrix}1\\[1pt]0\end{psmallmatrix}$ & $\bar\varphi^{\,0}$ & $+\tfrac12$ & $-1$ & $0$ \\
  & & $\begin{psmallmatrix}0\\[1pt]1\end{psmallmatrix}$ & -$\bar\varphi^{+}$ & $-\tfrac12$ & $-1$ & $-1$ \\
\bottomrule
\end{tabular}
\caption{First-generation fermion representations under $SU(2)_L\times U(1)_Y$, together with the conventional Higgs doublet and its conjugate. Here boldface on the quarks means each such field is a vector in $\mathbb{C}^3$. The pairs $(\mathbf u_L,\mathbf d_L)$, $(\nu_{eL},e_L)$, and $(\varphi^+,\varphi^0)$ transform as $SU(2)$ doublets. $Y$ is the hypercharge, $T_3$ is weak isospin, and $Q=T_3+\tfrac12Y$.}
\label{tab:fermion-reps}
\end{table}

Indeed, for $V_L, V_R$ irreducible, unitary, and non-isomorphic, $G$-invariant mass pairings $V_L\times V_R\rightarrow \mathbb{C}$ are necessarily trivial \cite[Theorem~7.6.11]{Hamilton_book}. The remedy is a \emph{Yukawa form}: a $G$-invariant trilinear map $\tau$, antilinear in $V_L$ and complex-linear in $W$ and $V_R$ (equivalently, an element of $\mathrm{Hom}_G(\bar V_L\otimes W\otimes V_R,\mathbb{C})$), where $W$ is the Higgs representation space or its conjugate, as the term requires. For the leptons, with $V_L=\mathbf{2}_{-1}$, $V_R=\mathbf{1}_{-2}$, $W=\mathbf{2}_{1}$ (cf.\ Table~\ref{tab:fermion-reps}), the Yukawa form and its extension to sections yield the gauge-invariant map
\begin{equation}\label{eq:T}
T(L_L,\varphi,e_R):=l_L^\dagger\phi\,l_R,
\end{equation}
where $l_L, \phi, l_R$ are the fibrewise representatives (see \cite[Ch.~8]{Hamilton_book} for the construction via local sections). But $T$ is not unique. The centraliser of the represented group, $H:=C_{GL(V_R)}(\rho_R(G))$ (the subgroup of $GL(V_R)$ commuting with the image of $G$), generates a family of invariant maps
\begin{equation}\label{eq:T'}
T'_h(L_L,\varphi,e_R):=l_L^\dagger\phi\,h\,l_R, \qquad h\in H.
\end{equation}
For $G=SU(2)\times U(1)$, the centraliser is $\mathbb{C}^\times$ and the ambiguity reduces to an overall complex factor, absorbable into the Yukawa coupling. Call each of the three one-generation field-content triples---$(\bar L_L,\varphi,e_R)$, $(\bar Q_L,\varphi,d_R)$, $(\bar Q_L,\tilde\varphi,u_R)$---a \emph{Yukawa channel}. For the Standard Model, this scalar exhausts each channel's ambiguity:
\begin{lem}[Multiplicity one]\label{lem:mult_one}
For each one-generation Yukawa channel, the space of $G$-invariant trilinear forms $\mathrm{Hom}_G(\bar V_L\otimes W\otimes V_R,\mathbb{C})$ is one-dimensional.
\end{lem}
\noindent The proof is a short Clebsch--Gordan count, given in Appendix~\ref{app:proofs}. Thus any two invariant Yukawa forms for a given channel differ by an overall scalar, and the family \eqref{eq:T'} exhausts the possibilities. This does not give the symmetry-first formulation two independent freedoms, a choice of form and then a choice of coupling.  After choosing a non-zero generator $T$ for the trilinear tensor $\mathcal{Y}$ that enters the Lagrangian, one may write $\mathcal{Y}=yT$, but
\be\label{eq:reciprocal}
(T,y)\sim(cT,c^{-1}y),
\qquad c\in\mathbb{C}^{\times}.
\ee
Representation theory therefore determines the invariant line, but it does not determine a preferred generator for this line. Changing the generator merely changes the numerical value assigned to $y$. But it is important to note that for more complicated groups and representations, invariant spaces need not be one-dimensional; different directions in such a space need not be absorbable into a single scalar coefficient.

\subsection{The geometry-first perspective}\label{sec:Yukawa2}

The geometry-first perspective treats Yukawa forms as natural operations between vector bundles: tensor products, canonical contractions, Hermitian pairings, and the map induced by the unit complex volume form. This lies in contrast to the symmetry-first description, where they are  equivariant maps on representation spaces.  Here the inputs are the fundamental bundle data~\eqref{eq:fund_bundle_data}. In particular, $E^1$ is fixed so that $(E^1)^{\otimes3}$ has hypercharge $+1$ in the units of Table~\ref{tab:fermion-reps}, and hence $E^1$ itself carries $+\tfrac13$. The internal bundle assignments are
\begin{align}
&\mathbf{L}_L\in\Gamma\big(E^2\otimes(E^{1*})^{\otimes3}\big), \qquad e_R\in\Gamma\big((E^{1*})^{\otimes6}\big),\qquad \mathbf{Q}_L\in\Gamma\big(E^3\otimes E^2\otimes E^1\big),\nonumber\\
&\mathbf{d}_R\in \Gamma\big(E^3 \otimes (E^{1*})^{\otimes 2}\big),\qquad \mathbf{u}_R\in\Gamma\big(E^3\otimes(E^1)^{\otimes4}\big),\qquad \boldsymbol{\varphi}\in\Gamma\big(E^2\otimes(E^1)^{\otimes3}\big),\label{eq:assignments}
\end{align}
with the spinor factors of Section~\ref{sec:Yukawa} left implicit. The classical gauge fields are represented by the compatible covariant derivatives $\nabla^1,\nabla^2,\nabla^3$ on the fundamental bundles (see \citealp{Gomes_internal,Gomes_AB}).

For these particular assignments, $\mathbf L_L$, $\mathbf Q_L$, and $\boldsymbol\varphi$ carry the rank-two factor $E^2$, whereas $e_R$, $\mathbf u_R$, and $\mathbf d_R$ do not. I use this bundle description from this point onward.\footnote{This is a statement about the matter bundles in~\eqref{eq:assignments}. But it is not a general definition of a trivial $SU(2)$ action: a trivial representation can also arise after tensor factors have been contracted.}

Before the Higgs section is specified, $\mathbf L_L$ and $\mathbf Q_L$ are each single tensor fields. Because $(E^1)^{\otimes3}$ is a line bundle, a nowhere-vanishing Higgs section canonically determines a complex line sub-bundle $\ell_H\subset E^2$. More explicitly, at each $x\in M$ the tensor $\boldsymbol\varphi_x$ defines a rank-one map
\[
\Phi_x:(E_x^{1*})^{\otimes3}\longrightarrow E_x^2,
\qquad
\alpha\longmapsto (\operatorname{id}_{E_x^2}\otimes\alpha)(\boldsymbol\varphi_x),
\]
and $\ell_{H,x}:=\operatorname{im}\Phi_x$. The Hermitian metric on $E^2$ then gives the orthogonal splitting
\[
E^2=\ell_H^\perp\oplus\ell_H.
\]
Let $P_H$ and $P_H^\perp$ denote the corresponding orthogonal projectors. The familiar left-handed fields are the projections of $\mathbf L_L$ and $\mathbf Q_L$ onto these two line sub-bundles:
\begin{align}
\boldsymbol e_L&:=\big(P_H\otimes\operatorname{id}_{(E^{1*})^{\otimes3}}\big)\mathbf L_L,
&\boldsymbol\nu_{eL}&:=\big(P_H^\perp\otimes\operatorname{id}_{(E^{1*})^{\otimes3}}\big)\mathbf L_L,\label{eq:QH1}\\
\mathbf d_L&:=\big(\operatorname{id}_{E^3}\otimes P_H\otimes\operatorname{id}_{E^1}\big)\mathbf Q_L,
&\mathbf u_L&:=\big(\operatorname{id}_{E^3}\otimes P_H^\perp\otimes\operatorname{id}_{E^1}\big)\mathbf Q_L.\label{eq:QH2}
\end{align}
In a local unit frame $e_H$ of $\ell_H$ (locally the $e_0^2$ of Section~\ref{sec:SU2}) we may write $\boldsymbol e_L=e_H\otimes e_L$ and, after choosing a colour frame, $\mathbf d_L^I=e_H\otimes d_L^I$. Thus the Higgs section defines the neutrino/charged-lepton and up/down decompositions. In symmetry-first language, it fixes the basis in which $T_3$ (defined only relative to such a basis; cf. footnote~\ref{ftnt:T3}) and the component labels of Table~\ref{tab:fermion-reps} are written. 

The two fields in each pair are therefore Higgs-relative components of one tensor field, not independently specified tensor fields.\footnote{\label{ftnt:table-misleading}Some textbooks note this (cf.\ \cite[p.~185]{Tong_SM}), though usually as a fact about unification `before symmetry breaking', rather than as a consequence of the Higgs-selected splitting in the broken theory.} The loop-mixing remark at the end of Section~\ref{sec:Higgs_field} can now be stated precisely: if $\nabla^2$ does not preserve $\ell_H\oplus\ell_H^\perp$, parallel transport around a closed loop can return $\boldsymbol e_L$ with a component in $\ell_H^\perp$, and hence partly as $\boldsymbol\nu_{eL}$.

Write
\[
E_L:=E^2\otimes(E^{1*})^{\otimes3},
\qquad
E_e:=(E^{1*})^{\otimes6}.
\]
Recalling that $E_H=E^2\otimes(E^1)^{\otimes3}$, there is a canonical isomorphism $E_H\otimes E_e\rightarrow E_L$, obtained by contracting the line-bundle factors. Hence the lepton Yukawa term is simply the Hermitian pairing on $E_L$:
\be\label{eq:Yukawa_L}
T_e(\mathbf L_L,\boldsymbol\varphi,e_R)
   =y_e\big\langle\mathbf L_L,\boldsymbol\varphi\otimes e_R\big\rangle_{E_L}
   =y_e(v+H)\,\bar e_L e_R,
\ee
where the Lorentz-spinor contraction remains implicit and $e_L$ denotes the coefficient of the $\ell_H$ component in the local frame above. Both sides of the first equality are frame-independent, whereas the last expression uses a Higgs-adapted frame.

The fibre structures therefore supply  a preferred normalised contraction in each one-dimensional invariant space of Lemma~\ref{lem:mult_one}. The scalar multiplying this contraction is the Yukawa coupling. Geometry thus gives a canonical separation between the form of the interaction and its strength; an explanatory advantage relative to the symmetry-first version of the Yukawa form, but it is not enough for a reduction in free parameters: the coupling itself, and with three generations the Yukawa matrix, remains free. 

So far, the contrast with the symmetry-first formulation  stems from the fact that, as we saw, representation theory doesn't fix a unique Yukawa map, for one may relabel the representation data according to Lemma \ref{lem:mult_one}. But, in the symmetry-first description, does representation theory allow a wider degeneracy in the Yukawa map, a greater freedom than could be absorbed by re-adjusting  coupling constants? 

 To answer, write $G_W\subseteq GL(W)$ for the image of $G=SU(3)\times SU(2)\times U(1)$ on a Standard Model matter fibre $W$. The insertions of~\eqref{eq:T'} lie in the centraliser $C_{GL(W)}(G_W)$. However, the most general complex-linear re-identifications available to the symmetry-first formulation lie in the normaliser $N_{GL(W)}(G_W)$, whose elements intertwine the representation after re-identifying the group by $\operatorname{Ad}_u$ (cf. \citep{Gomes_nonequiv}). But we have the following: %This could mean there is more freedom in the symmetry-first description of a Yukawa mechanism than could be absorbed by coupling constants. However, this is shown not to be the case by  the following Proposition:

\begin{prop}[The Standard Model's Yukawa ambiguity is purely central]\label{prop:central}
For every matter fibre $W$ of Table~\ref{tab:fermion-reps}, $N_{GL(W)}(G_W)=\mathbb{C}^{\times}\!\cdot G_W=\mathbb{R}^{>0}\!\cdot G_W$.\footnote{The fibres of Table~\ref{tab:fermion-reps} are single-generation. Including a flavour factor $F\cong\mathbb{C}^3$ enlarges the commutant by $GL(F)$: the flavour-basis transformations. The Yukawa matrices are maps between distinct flavour spaces, and their physical content (masses and CKM mixing) is what survives quotienting by those basis changes, as treated below.} Consequently every symmetry-first re-identification acts on the Yukawa form~\eqref{eq:T} by an overall scalar: relabelling the group and representation yields nothing beyond the centraliser freedom of~\eqref{eq:T'}.
\end{prop}

Proposition~\ref{prop:central}, proved in Appendix~\ref{app:proofs}, is the fibre-level counterpart of the normaliser analysis in \citet{Gomes_nonequiv}. For the Standard Model's complex matter fibres, the excess symmetry-side self-identifications change only the generator chosen for the invariant Yukawa term, with the inverse change in its coefficient, as described in \eqref{eq:reciprocal}. They introduce no further physical freedom beyond one coefficient for each one-generation invariant contraction.

As noted below~\eqref{eq:assignments}, only the left-handed matter bundles carry an $E^2$ factor (that is the geometric expression of the Standard Model's chirality). Neutrinos remain massless in the minimal matter content, but that's not because $\boldsymbol\nu_{eL}$ lies in $\ell_H^\perp$ (which it does by definition). The reason is that no right-handed neutrino field has been included. Were such a field added, a Dirac neutrino coupling could use the conjugate Higgs field constructed below, just as the up-type quark coupling does.\footnote{With only the charged-lepton Yukawa map, independent changes of frame in the left- and right-handed lepton multiplicity spaces diagonalise that map with no observable residue. In the minimal Standard Model, without a neutrino mass map, there is therefore no leptonic analogue of the CKM matrix.}

In conventional component notation the required conjugate Higgs field is $\tilde\phi=i\sigma_2\bar\phi$, the $\varphi_c$ of Table~\ref{tab:fermion-reps}. Its frame-free construction uses the unit complex volume form $\epsilon:=\epsilon_2$ on $E^2$.\footnote{Under a unitary automorphism $A$ of the rank-two fibre, $A^*\epsilon=\det(A)\epsilon$, with $|\det(A)|=1$. Thus $\operatorname{Stab}_{U(2)}(\epsilon)=SU(2)$: fixing $\epsilon$ is equivalent to requiring $\det(A)=1$.\label{ftnt:orient}} The electroweak mass calculation of Section~\ref{sec:SU2} used only the Hermitian structure induced on $E_H$; as anticipated in the Remark opening Section~\ref{sec:Higgs}, the up-type Yukawa contraction is the first point at which the factorisation $E_H=E^2\otimes(E^1)^{\otimes3}$ and the volume form on its rank-two factor are essential.

In order to find the frame-free construction of the conjugate Higgs, we recall that there are now two identifications of $E^2$ with its dual. The Riesz map $J:E^2\rightarrow E^{2*}$, $\xi\mapsto\langle\xi,\cdot\rangle_2$, is conjugate-linear, while $\epsilon^\flat:E^2\rightarrow E^{2*}$, $\xi\mapsto\epsilon(\xi,\cdot)$, is complex-linear and has inverse $\epsilon^\#$. Their composition is the canonical quaternionic structure on the rank-two Hermitian fibre,
\be\label{eq:Cdef}
C:=\epsilon^\#\circ J:\;E^2\longrightarrow E^2.
\ee
In an $\epsilon$-adapted unitary frame, $C\xi=i\sigma_2\bar\xi$. Intrinsically,
\[
C^2=-\mathbb{1},\qquad
\langle C\xi,C\eta\rangle_2=\langle\eta,\xi\rangle_2,\qquad
\langle C\xi,\xi\rangle_2=0,\qquad
CA=AC\quad(A\in SU(2)),
\]
so $C\xi$ is the Hermitian-orthogonal partner of $\xi$, of the same norm, with phase fixed by $\epsilon$.

Because $C$ is conjugate-linear, $C\otimes\operatorname{id}$ does not define a map on the complex tensor product $E_H=E^2\otimes(E^1)^{\otimes3}$: the line factor must be conjugate-dualised at the same time.\footnote{The two presentations $(\lambda v)\otimes\ell=v\otimes(\lambda\ell)$ of one tensor would otherwise be sent to $\bar\lambda C(v)\otimes\ell$ and $\lambda C(v)\otimes\ell$. On decomposable elements, the well-defined map is $\widetilde C(v\otimes\ell)=C(v)\otimes J_L(\ell)$, with $J_L$ the Riesz map of the line factor. Appendix~\ref{app:proofs} gives its frame expression and transformation law.} Applying the Riesz map of the full Higgs bundle and then $\epsilon^\#$ on the rank-two factor gives
\be\label{eq:Ctilde}
\widetilde C:=(\epsilon^\#\otimes\operatorname{id})\circ J_H:
E^2\otimes(E^1)^{\otimes3}\longrightarrow
E^2\otimes(E^{1*})^{\otimes3}=:E_{\widetilde H}.
\ee
The section
\[
\boldsymbol\varphi_c:=\widetilde C(\boldsymbol\varphi)\in\Gamma(E_{\widetilde H})
\]
is the bundle-level conjugate Higgs field: its rank-two factor remains $E^2$, while its line-bundle factor is replaced by its dual.

The right-handed quark fields take values in the inequivalent bundles
\[
E_u:=E^3\otimes(E^1)^{\otimes4},
\qquad
E_d:=E^3\otimes(E^{1*})^{\otimes2},
\]
neither of which contains an $E^2$ factor. They are therefore not the two Higgs-relative projections of a single rank-two-valued field.\footnote{\label{ftnt:RH_split}They can of course be placed in the direct sum $E_u\oplus E_d$, but this does not turn them into projections of one field along a common $E^2$ factor. Had the two fields instead arisen from one $E^2$-valued field, their Yukawa terms would reduce to the contractions along $\ell_H$ and $\ell_H^\perp$. %Minimal $SU(5)$ retains the split: the up- and down-type Yukawas remain the distinct contractions $\mathbf{10}\cdot\mathbf{10}\cdot\mathbf{5}_H$ and $\mathbf{10}\cdot\bar{\mathbf{5}}\cdot\bar{\mathbf{5}}_H$.
} Tensoring $\mathbf d_R$ with $\boldsymbol\varphi$, and $\mathbf u_R$ with $\boldsymbol\varphi_c$ while contracting the $E^1$ factors aligns both with the internal bundle
\[
E_Q:=E^3\otimes E^2\otimes E^1
\]
of $\mathbf Q_L$. Indeed, contraction of the line factors gives the canonical isomorphisms
\be\label{eq:EQiso}
E_H\otimes E_d\cong E_Q\cong E_{\widetilde H}\otimes E_u.
\ee
The quark Yukawa term is consequently one Hermitian pairing on $E_Q$. With $i,j=1,2,3$ indexing generations,
\be\label{eq:Yukawa_Q}
T_q=
\Big\langle\mathbf Q_L^i,
\boldsymbol\varphi\otimes Y^d_{ij}\mathbf d_R^j
+\boldsymbol\varphi_c\otimes Y^u_{ij}\mathbf u_R^j
\Big\rangle_{E_Q},
\ee
the geometric form of the standard definition (cf.~\cite[Lemma~8.8.4]{Hamilton_book}). In a Higgs-adapted frame, where $\varphi=(0,H+v)^T$ and $\varphi_c=(H+v,0)^T$,
\be\label{eq:Yukawa_Q2}
T_q=(H+v)\left(Y^d_{ij}\,\bar d_L^{\,Ii}d_R^{Ij}+Y^u_{ij}\,\bar u_L^{\,Ii}u_R^{Ij}\right),
\ee
summing over colour $I$.

\subsection{The CKM matrix}
What, then, is the status of the generation index? The three fields
\(\mathbf{Q}^1_L,\mathbf{Q}^2_L,\mathbf{Q}^3_L\) have the same Lorentz
and internal type. Their threefold direct sum may therefore be written as
\[
E_Q^{\oplus 3}\cong E_Q\otimes F_Q,
\]
where \(F_Q\) is a three-dimensional Hermitian multiplicity space. Choosing
an orthonormal basis \((f_1,f_2,f_3)\) of \(F_Q\) gives
\[
\mathbf{Q}_L=\sum_i\mathbf{Q}^i_L\otimes f_i.
\]
The generation labels are thus basis indices in \(F_Q\), not labels of
different bundles. A different orthonormal basis presents the same field
through different linear combinations of its components. The bundle \(E_Q\) thus 
encodes the field's internal type, while \(F_Q\) encodes the multiplicity of this type (i.e.  multiplicity three). In particular, differences of mass are not encoded
in \(E_Q\).

The same construction applies independently to each quark field:
\[
\mathbf{Q}_L\in
\Gamma(S_L\!\otimes\!E_Q\!\otimes\!F_Q),\qquad
\mathbf{u}_R\in
\Gamma(S_R\!\otimes\!E_u\!\otimes\!F_u),\qquad
\mathbf{d}_R\in
\Gamma(S_R\!\otimes\!E_d\!\otimes\!F_d).
\]
The spaces \(F_Q,F_u,F_d\) are each isomorphic to \(\mathbb{C}^3\), but of course 
there is no canonical identification between them. Each field type has its
own multiplicity space and hence its own frame choices. So these factors don't
introduce any additional field content but they rewrite the corresponding
threefold direct sums, with the same covariant derivative acting on each
copy.\footnote{If \(F\) is regarded as the trivial bundle \(M\times F\), then
under \(E^{\oplus3}\cong E\otimes F\) the diagonal derivative
\(\nabla^{\oplus3}\) corresponds to the tensor-product derivative induced by
\(\nabla\) and the fixed flat derivative on \(M\times F\). No independent
flavour connection is thereby introduced.}

The Yukawa maps are fixed linear maps between the multiplicity spaces,
\[
Y_u\in\operatorname{Hom}(F_u,F_Q),\qquad
Y_d\in\operatorname{Hom}(F_d,F_Q).
\]
In invariant terms, \eqref{eq:Yukawa_Q} pairs \(\mathbf{Q}_L\) with
\[
\boldsymbol{\varphi}\otimes Y_d\mathbf{d}_R
   +\boldsymbol{\varphi}_c\otimes Y_u\mathbf{u}_R
\]
over \(E_Q\otimes F_Q\). Choosing frames turns \(Y_u\) and \(Y_d\) into
matrices, but the maps themselves are independent of that choice. The
freedom to change flavour frames therefore remains, and the Yukawa maps
supply a basis-independent spectral structure on the flavour spaces.

Since \(Y_u\) maps \(F_u\) into the different space \(F_Q\), an equation
such as \(Y_ur=\lambda r\) is not well-defined. But we can use the relevant eigenvalue
problem on the left-handed flavour space, defined by the positive
self-adjoint operator
\[
A_u:=Y_uY_u^\dagger\in\operatorname{End}(F_Q),
\]
and likewise
\[
A_d:=Y_dY_d^\dagger\in\operatorname{End}(F_Q).
\]
After the Higgs takes its vacuum value \(v\), the corresponding
left-handed mass-squared operators are \(v^2A_u\) and \(v^2A_d\). Now choose 
orthonormal eigenbases satisfying
\[
A_u\ell^u_i=(y^u_i)^2\ell^u_i,\qquad
A_d\ell^d_j=(y^d_j)^2\ell^d_j,
\]
where \(y^u_i,y^d_j\geq0\). The numbers \(y^u_i\) and \(y^d_j\) are the
singular values of the Yukawa maps: the non-negative square roots of the
eigenvalues of \(Y_uY_u^\dagger\) and \(Y_dY_d^\dagger\).

The singular-value decomposition also supplies orthonormal bases
\((r^u_i)\) of \(F_u\) and \((r^d_j)\) of \(F_d\) such that
\be\label{eq:SVD}
Y_u r^u_i=y^u_i\ell^u_i,\qquad
Y_d r^d_j=y^d_j\ell^d_j.
\ee
Equivalently, the \(r^u_i\) and \(r^d_j\) are eigenvectors of
\(Y_u^\dagger Y_u\) and \(Y_d^\dagger Y_d\), respectively. In these
singular-vector bases, \eqref{eq:Yukawa_Q2} becomes a sum of mass terms
with
\[
m^u_i=v\,y^u_i,\qquad m^d_j=v\,y^d_j.
\]
Thus a non-degenerate mass eigenstate pairs a right-handed one-dimensional vector space, such as
\(\mathbb{C}r^u_i\subset F_u\), with a left-handed one
\(\mathbb{C}\ell^u_i\subset F_Q\).\footnote{If a singular value is degenerate, the
Yukawa map selects the corresponding higher-dimensional eigenspaces. In either case, different masses require no difference
in the underlying bundles.}

To express the same result in matrix form, choose orthonormal reference
frames in \(F_Q,F_u\), and \(F_d\). Let \(U_{uL}\) be the unitary matrix
whose columns are the coordinate vectors of the left singular vectors
\((\ell^u_i)\) in \(F_Q\), and let \(U_{dL}\) be defined similarly from
\((\ell^d_j)\). Let \(U_{uR}\) and \(U_{dR}\) be the unitary matrices whose
columns are the right singular vectors \((r^u_i)\) in \(F_u\) and
\((r^d_j)\) in \(F_d\), respectively. The matrices representing the Yukawa
maps then take the form
\[
Y_u=U_{uL}D_uU_{uR}^\dagger,\qquad
Y_d=U_{dL}D_dU_{dR}^\dagger,
\]
where
\[
D_u=\operatorname{diag}(y^u_1,y^u_2,y^u_3),\qquad
D_d=\operatorname{diag}(y^d_1,y^d_2,y^d_3).
\]

The important point for mixing is that \(A_u\) and \(A_d\) both act on the
same left-handed flavour space \(F_Q\). They therefore determine two
generally different orthonormal eigenframes of that space,
\((\ell^u_i)\) and \((\ell^d_j)\). A single common mass frame exists
precisely when
\[
[A_u,A_d]=0.
\]
But in general they do not commute, and the relative unitary between their
eigenframes is the CKM matrix:
\be\label{eq:CKM}
(V_{\mathrm{CKM}})_{ij}
   =\langle\ell^u_i,\ell^d_j\rangle_{F_Q},
\qquad\text{i.e.}\qquad
V_{\mathrm{CKM}}=U_{uL}^\dagger U_{dL}.
\ee
%This matrix is defined up to the usual permutations and rephasings of non-degenerate mass eigenstates, together with unitary rotations within any degenerate mass eigenspaces.

Now, the comparison between the two left-handed mass frames is canonical because the up- and down-type fields are the two Higgs-relative projections of the single tensor field $\mathbf Q_L$. Under the splitting $E^2=\ell_H^\perp\oplus\ell_H$, both inherit the same  $F_Q$. If they were unrelated fields with independent multiplicity spaces, their mass operators could be diagonalised independently and no relative unitary would be defined. Here, however, the off-diagonal part of $\nabla^2$ (relative to the Higgs-selected splitting) carries one component into the other and acts as $\operatorname{id}_{F_Q}$ on flavour. One may diagonalise the two mass operators in their respective eigenframes, but their mismatch then appears in this connection term: written between the up- and down-type mass frames, the identity on $F_Q$ has matrix
\[
U_{uL}^\dagger\,\operatorname{id}_{F_Q}\,U_{dL}
   =U_{uL}^\dagger U_{dL}
   =V_{\mathrm{CKM}}.
\]

There is no analogous comparison of the right-handed mass frames. The fields $\mathbf u_R$ and $\mathbf d_R$ take values in $E_u$ and $E_d$ and have no $E^2$ factor (Section~\ref{sec:Yukawa2}), so they are not the two projections of one field along a Higgs-selected splitting, and thus no off-diagonal component of $\nabla^2$ relates them. Their flavour frames can consequently be changed independently which  is why $U_{uR}$ and $U_{dR}$ enter the singular-value decompositions of the Yukawa maps but not the CKM matrix.

\section{Scope, constraints, and explanatory limits}
\label{sec:defense}

The geometry-first picture binds gauge groups, fundamental fibres and matter representations together more tightly than the principal-bundle picture does. Some principal-bundle presentations fall outside the resulting tensorial construction altogether. Others remain formally available, but on primitive structure of diminishing geometrical motivation; for $E_8$ the explanatory gain is spent.

In Section \ref{sec:intro} I claimed that the principal-bundle picture tolerates a slack between symmetry and geometry that the geometry-first picture does not tolerate. It is time to make good on that claim. The slack is this: in the PFB-POV, the group $G$, the representation $\rho$, and the fibre $V$ can be chosen independently, subject only to mutual consistency. The geometry-first picture eliminates this freedom.\footnote{Throughout this section, $\mathsf{Aut}(V)$ means the group of automorphisms preserving whatever fibre structure is in play (e.g.\ Hermitian form, and where specified a unit volume form).} The argument proceeds in two stages: I exhibit the slack through examples (Section~\ref{sec:slack}), then examine three further points of divergence between the pictures (Section~\ref{sec:E6}). That the VB-POV's constraints are assumptions physicists already tacitly make is taken up at the close of Section~\ref{sec:conclusions}.

\subsection{Slack between symmetry and geometry: examples and lessons}
\label{sec:slack}

Recall from Section~\ref{sec:PFB} that an associated bundle $E = P \times_\rho V$ requires only that $\rho(G)$ preserve the structure of $V$:
\be\label{eq:subset}
\rho(G)\subseteq \mathsf{Aut}(V)\subseteq GL(V).
\ee
The geometry-first picture demands more of any bundle taken as \emph{fundamental} for a gauge factor: the factor must be identified, by $\rho$ itself, with the full automorphism group of the declared fibre structure,
\be\label{eq:aut}
\rho:G\rightarrow\mathsf{Aut}(V),
\qquad\text{with}\qquad
\ker\rho=\{\mathbb{1}\}
\quad\text{and}\quad
\rho(G)=\mathsf{Aut}(V).
\ee
The restriction to fundamental bundles is important because derived bundles carry whatever actions the tensorial constructions induce, and these may be non-faithful (the centre acts trivially on the adjoint bundle) or fail to exhaust the derived fibre's automorphisms (cf.\ Section~\ref{sec:Yukawa2}). Equation~\eqref{eq:aut} conjoins two conditions---one on the kernel, one on the image---which can fail independently. The PFB-POV tolerates both failures, which are not, however, on the same footing. Three examples make this concrete.

\emph{Example 1 (trivial kernel, proper image).}
Take a Hermitian vector bundle $(E\to M,\langle\,\cdot\,,\,\cdot\,\rangle)$ with typical fibre $\mathbb{C}^n$ and a compatible covariant derivative $\nabla$, so that $\mathsf{Aut}(E_x)\simeq U(n)$. The PFB-POV may nonetheless posit $G=SU(n)$ in the defining representation. The action is faithful, but
\be
\rho(G)=SU(n)\subsetneq U(n)=\mathsf{Aut}(V).
\ee
Thus the posited group preserves more structure---a complex volume form---than the fibre structure contains. I call this a \emph{slack world}. The slack has a  physical character: the connection $\varpi$ carries observables independently of the matter content, and the Yang--Mills equations are valued in $\mathfrak{su}(n)$ rather than $\mathfrak{u}(n)$. The $G=SU(n)$ and $G=U(n)$ worlds are different theories although their matter content is identical.\footnote{Parallel transport around closed curves at $x\in M$ generates the holonomy group $\mathsf{Hol}(\nabla)\subseteq\mathsf{Aut}(V)$, and under natural assumptions a principal bundle $(P',M,G')$ exists with $G'\simeq\mathsf{Hol}(\nabla)$ (cf.\ \cite[Theo.~17.11]{Michor2008}). This does not close the gap: holonomy belongs to a particular connection, whereas the slack concerns the group posited by the theory.}

The VB-POV forbids this combination. Each of its gauge factors is $\mathsf{Aut}(E_x)$, so  the Hermitian fibre yields the $U(n)$ theory, and the $SU(n)$ theory is obtained only by enriching the fundamental structure with a unit complex volume form $\epsilon$, which trivialises the determinant line $\det(E)=\Lambda^nE$ and reduces $U(n)$ to $SU(n)$. Either way the group is not posited but fixed by the fibre geometry. Everything else is then derived from $(E,\nabla)$: for instance, the adjoint bundle $\mathfrak{su}(E)$ of traceless anti-Hermitian endomorphisms $\mathsf{End}(E)\simeq E^*\otimes E$, whose one-form sections parametrise the affine space of compatible derivatives (what are called ``gluon fields'' correspond, in the VB-POV, to $\nabla$ itself); and the matter bundles $E$, $E^*$, $\mathrm{Sym}^k(E)$, $\Lambda^k E$, each with its induced derivative.

\emph{Example 2 (continuous kernel).}
Let $V=\mathbb{C}^2$ carry a Hermitian inner product and a unit complex volume form, so that $\mathsf{Aut}(V)\simeq SU(2)$, and take $G=\mathrm{Spin}(4)\cong SU(2)_L\times SU(2)_R$, acting through the left-handed chiral representation, $\rho(g_L,g_R)=g_L$. The image now exhausts $\mathsf{Aut}(V)$, but the kernel is $SU(2)_R$. This can be interpreted as a vacuum sector rather than as slack. In more detail, the VB-POV can accommodate this case by giving each factor its own fundamental bundle: the idle one carries no matter, and its gauge sector satisfies the vacuum Yang--Mills equations (cf.\ Equation~\eqref{eq:source} in Appendix~\ref{app:PFB}). This works because $SU(2)_R$ is a direct factor.\footnote{A kernel is always normal, but normality is not enough, and neither is a split extension $1\to K\to G\to G/K\to 1$, which yields only a semidirect product; independent fundamental bundles compose as a \emph{direct} product, so only a direct factor detaches as its own bundle. See \citet{Gomes_nonequiv} for the classification, and for the Standard Model's diagonal $\mathbb{Z}_6$, a kernel that is not a direct factor.} The degenerate limit is a trivial representation, $\rho(G)=\mathbb{1}$: both conditions in \eqref{eq:aut} then fail, and $G$ cannot be reconstructed from its action on $V$ at all. Either way, a \emph{single} matter bundle cannot recover the full gauge group. This point will resurface in Section~\ref{sec:composite}.

\emph{Example 3 (discrete kernel).}
Let $V=\mathbb{R}^{2n}$ carry a Euclidean inner product and an orientation, so that $\mathsf{Aut}(V)= SO(2n)$, and let $G=\mathrm{Spin}(2n)$ act through the vector representation $\rho:\mathrm{Spin}(2n)\to SO(2n)$, with kernel $\{\pm1\}$. Since $\ker(\rho_*)=0$, the local field equations are those of an $SO(2n)$ theory; the difference is purely global, concerning the topology of the gauge group and hence which principal bundles are topologically distinct. For the VB-POV the distinction is crucial. For we obtain a global group from  $\mathsf{Aut}(V,\mathrm{struct})$, not a Lie algebra, and the global form is wholly fixed by the choice of fundamental fibre and its structures. This fibre yields $SO(2n)$, and recovering $\mathrm{Spin}(2n)$ would require fundamental structure on which the central element acts non-trivially \citep{Gomes_globalform}.

In sum, the two conditions of \eqref{eq:aut} can fail independently, and the PFB-POV tolerates every combination. The VB-POV rules on each case by a single principle: each factor of the gauge group \emph{is} the automorphism group of some fundamental bundle. It forbids the slack world of Example~1: no fundamental bundle has $SU(n)$ as the automorphism group of a merely Hermitian fibre. It accommodates the vacuum sector of Example~2 by splitting the group across two fundamental bundles, one idle (possible because the kernel is a direct factor). And, as to Example~3, where the local equations leave the choice of global form of the group given the algebra, the fundamental fibre delivers a definite global form. The deeper non-equivalences between the two frameworks---the diagonal-kernel obstruction, the non-recovery of tensorial genealogy, the strict inclusion of theory-spaces, and a difference of symmetry content---are analysed in \citet{Gomes_nonequiv}.

\subsection{Fundamental obstacles to equivalence}\label{sec:E6}

The preceding examples stayed within the classical families. The differences between the two pictures run deeper: they explain charge quantisation differently; for some gauge groups the geometry-first picture reaches genuine limits; and matter content alone does not determine the group.

\paragraph{Quantisation of charge.}\label{par:charge}
The geometry-first picture enforces charge quantisation by construction. All charged fields arise as tensor powers $E^{\otimes n}$ and duals of a fundamental line bundle $E$; the integer label $n$ counts fibre copies, so charges form a discrete lattice. This contrasts with the symmetry-first picture, where this kind of discreteness is traced to the topology of the group $U(1)$; or more specifically, to its compactness. The tensorial construction, but not the symmetry-first one, enforces discreteness even when $\mathsf{Aut}(V)\simeq\mathbb{C}^\times$ is non-compact. 
The two explanations also differ in modal commitment. Having fixed the Lagrangian, the symmetry-first discreteness is conditional on a \emph{choice} of the compact global form of the group, for the Lagrangian sees only $\mathfrak{u}(1)\cong\mathbb{R}$, and thus nothing local is evidence for $U(1)$ over the additive group $\mathbb{R}$. It is only the observed commensurability of the known charges that supports the  choice of $U(1)$. Indeed, a newly discovered incommensurate charge would simply be assigned a character of $\mathbb{R}$ in the symmetry-first approach. In contrast, geometry-first discreteness is not conditional on the spectrum: every matter bundle derived from the fundamental line carries an integer charge, so an incommensurate discovery  would falsify the VB-POV  itself.\footnote{Division-algebraic reconstructions find integer charges by a different route: \citet{Furey_charge} obtains electric charge as the eigenvalue of a number operator; here the integer counts tensor factors.}

Thus I find this explanation of discreteness considerably more satisfying than the topological one; a fuller treatment, including the modal contrast, appears in \citet{Gomes_nonequiv}.

\paragraph{Exceptional Lie groups.}
A different kind of obstacle arises with the exceptional Lie groups. The issue is not that the VB-POV's machinery fails (one can always build associated vector bundles given any representation of any compact group), but that the \emph{explanatory direction} fails, or rather, becomes strained.

Recall that the VB-POV fixes a fibre $V$ endowed with invariant geometric structure and defines $G=\mathsf{Aut}(V,\text{structure})\subset GL(V)$. For the classical families, the structure determines a unique group whose action both preserves and exhausts the geometry of $V$. For the exceptional families, the correspondence ranges from nearly as clean to circular.

$G_2$ is at the clean end of the spectrum. It arises as $\mathsf{Aut}(\mathbb{R}^7, \varphi)$, where $\varphi\in\Lambda^3(\mathbb{R}^7)^*$ is a generic (stable) three-form. % of definite type---stable three-forms in dimension seven fall into two open orbits, and the one whose induced bilinear form is definite yields the compact form. 
 The three-form determines an inner product and an orientation automatically (so $G_2\subset SO(7)$), as well as a cross product $\times:\mathbb{R}^7\times\mathbb{R}^7\rightarrow\mathbb{R}^7$ via $g(u\times v, w)=\varphi(u,v,w)$ (equivalently, the imaginary octonion multiplication).\footnote{$G_2$-geometry is a mainstream topic in differential geometry: the holonomy classification of Berger lists $G_2$ as one of the possible holonomy groups of irreducible non-symmetric Riemannian manifolds (in dimension $7$), characterised precisely by the existence of a parallel three-form.} The fibre structure would be $(V, \varphi)$ with $V\simeq\mathbb{R}^7$, matter fields would be ordinary sections of the associated rank-$7$ bundle, and all higher representations would be built tensorially. A three-form is a standard tensor, and the VB-POV is applicable in exactly its usual pattern.%---indeed more economically than for $SU(n)$, since a single tensor does the work of both an inner product and a volume form.

$E_6$, $F_4$, and $E_7$ admit analogous realisations, but they all depend on structures rooted in the exceptional Jordan algebra $J_3(\mathbb{O})$. $E_6$ arises as $\mathsf{Aut}(\mathbb{C}^{27}, \langle\cdot,\cdot\rangle, C)$, where $C\in\mathsf{Sym}^3(V^*)$ is the cubic form given by the determinant of $J_3(\mathbb{O})$. $F_4$ arises as $\mathsf{Aut}(J_3(\mathbb{O}), \circ)$, the automorphism group of the exceptional Jordan algebra of Hermitian $3\times 3$ octonionic matrices.\footnote{Equivalently, on the $26$-dimensional traceless subspace $J_3(\mathbb{O})_0$, it is the group preserving the quadratic and cubic trace forms.} Non-compact real forms of $E_7$ arise as $\mathsf{Aut}(\mathbb{R}^{56}, \omega, q)$, where $\omega\in\Lambda^2(V^*)$ is a symplectic form and $q\in\mathsf{Sym}^4(V^*)$ is a quartic invariant (also traceable to $J_3(\mathbb{O})$ via the Freudenthal construction).\footnote{The compact form acts quaternionically: its $\mathbf{56}$ is pseudo-real, with no real $56$-dimensional model. %It sits instead inside $\mathsf{Aut}(\mathbb{C}^{56},\omega,q)$ as the subgroup preserving a compatible positive Hermitian form.
}
 In each case, strictly speaking, the VB machinery is applicable: sections of the associated bundles are ordinary vector-valued fields, and all representations are built tensorially from the fundamental one. But the  claim of conceptual priority now loses support, as there is less {geometrical motivation} for the group. Inner products preserve norms and volume forms encode orientations; a cubic form on $\mathbb{C}^{27}$, a Jordan product on the $27$-dimensional Albert algebra, and a quartic form on a $56$-dimensional space all trace back to the same algebraic source: the exceptional Jordan algebra. This is a symmetry-first structure that would have to be postulated as primitive fibre structure without comparable geometric rationale. 

In this spectrum, $E_8$ is the worst case in  terms of geometrical motivation in my sense. Its minimal faithful representation is the \emph{adjoint}, of dimension $248$. Again, one can write down a vector bundle whose automorphism group is $E_8$:
\be
E = (\mathfrak{e}_8, \kappa, [\cdot,\cdot]),
\ee
where $\mathfrak{e}_8$ carries the Killing form $\kappa$ and the Lie bracket $[\cdot,\cdot]\in\Lambda^2(V^*)\otimes V$. The Lie bracket is a tensor and the Jacobi identity is an algebraic constraint. Thus, although this structure might be called `exotic', that is not the source of the obstruction. The obstruction is a certain kind of \emph{circularity}. For here there is no small ``matter'' representation analogous to the fundamental representations of the classical groups: the ``fundamental bundle'' is the gauge algebra itself. The VB-POV's claim of conceptual priority---``start with matter, recover the gauge group''---loses support when the matter space from which the group is to emerge is already the group's own Lie algebra. Again, strictly speaking the VB machinery still runs, but now without any explanatory benefit. These cases are not  geometry-first. 

|Moreover, the algebraic source that appears in these cases is also where other several current reconstructions of the Standard Model's  group begin \citep{DVTodorov_Jordan,Boyle_triality}. One such reconstruction exhibits the group as the stabiliser of additional structure inside $F_4=\mathsf{Aut}(J_3(\mathbb{O}),\circ)$ \citep{BaezSchwahn_Jordan}.  Again, from the present standpoint these constructions are `symmetry-first' in that the group is recovered as a proper subgroup of the automorphism group of a fibre whose primitive structure is Jordan-theoretic, not as the full automorphism group of a fibre carrying standard geometric structure.

To summarise: the five exceptional families arrange themselves along a gradient.  $G_2$ fits the VB-POV cleanly, with a single standard tensor (a three-form) playing the role that inner products and volume forms play for the classical groups. $E_6$, $F_4$, and $E_7$ fit formally, but all rely on structures descended from the exceptional Jordan algebra $J_3(\mathbb{O})$, which already counts as symmetry structure. And $E_8$ is circular: the matter space \emph{is} the symmetry structure.

This is a genuine limitation of the geometry-first picture. The VB-POV is most naturally suited to theories whose gauge groups are classical or admit low-dimensional faithful representations carrying standard geometric structures, which is precisely what the Standard Model provides.

\paragraph{Matter content.}\label{sec:composite}

One might hope that the mismatch between group and geometry disappears when the full matter content is taken into account: that the Standard Model’s many particle species, taken together, pin down $G$. \citet{Gomes_nonequiv} shows that they do not: no single matter bundle sees the full group; nothing intrinsic to a composite fibre such as the left-handed quark's $\mathbb{C}^3\otimes\mathbb{C}^2\otimes\mathbb{C}$ singles out the tensor decomposition from which the product structure of $G$ derives; and there can be intertwining obstructions, such as the diagonal $\mathbb{Z}_6$ that acts trivially on every Standard Model matter fibre, to recovering the product form from the matter bundles taken together.

 %As I have provided it here, the formulation is available only as an alternative to the gauge theories of linear groups. 

\section{Conclusions}\label{sec:conclusions}
This paper has made two kinds of claim: claims about \emph{explanatory structure} (the Higgs and Yukawa reformulations of Sections~\ref{sec:Higgs}--\ref{sec:Yukawa}) and claims about \emph{constraints and explanatory reach} (the slack diagnosis and exceptional-group discussion of Section~\ref{sec:defense}).

Regarding explanatory structure: in the geometry-first picture, the geometry of the spaces where matter lives has conceptual priority. Here, each gauge factor can arise solely as the automorphism group of a fundamental bundle. The formulation posits a different set of primitive objects, explains familiar mechanisms without invoking symmetry postulates, and structures the relationship between symmetry and geometry more tightly than the principal-bundle formulation would require. 

Not surprisingly, some explanations for known mechanisms differ in the two formulations. For the Higgs mechanism (Section~\ref{sec:Higgs}), mass acquisition reduces to extrinsic curvature: the shape operator of the sub-bundle singled out by the Higgs vacuum controls which components of the affine structure acquire mass, with no appeal to Goldstone modes, stabilisers, or Killing forms. The standard electroweak spectrum, Weinberg angle included, falls out of the shape operator explicitly (Section~\ref{sec:SU2}). Moreover,  the geometry-first picture does not merely offer a \emph{different} explanation of the Higgs mechanism; it dissolves a conceptual tension in the standard one, the equivocation on the status of $w_o$ analysed in Section~\ref{sec:Higgs_dyn}.

For the Yukawa mechanism (Section~\ref{sec:Yukawa}), fermion mass terms become fibrewise contractions built from the inner products and volume forms of the fundamental bundles. The coupling constant is thereby defined univocally, in contrast to the textbook presentation, where it can absorb the centraliser ambiguities in the definition of the Yukawa map (Proposition~\ref{prop:central}). The up-type Yukawa contraction depends essentially on the unit volume form of the weak fibre, whose stabiliser in $U(2)$ is $SU(2)$. The up- and down-type mass operators act on a single space, and the mismatch of their eigenframes is the CKM matrix (Section~\ref{sec:Yukawa}). For $\mathbb{C}^3$, the analogous mechanism, which would use the colour volume form and so select $SU(3)$, remains unclear.\footnote{There is a canonical isomorphism $U(3)\simeq (SU(3)\times U(1))/\mathbb{Z}_3$, and the Standard Model's $U(1)$ representations do realise it: in the units of Table~\ref{tab:fermion-reps}, every field's $E^1$ exponent is congruent mod~3 to its number of $E^3$ indices (duals counting negatively), which is exactly the condition for the matter to descend to $U(3)\times SU(2)$, one of the four admissible global forms of the Standard Model group \citep{Gomes_globalform}; the four quotients are distinguished physically by their admissible line operators \citep{Tong_line}. Baryon couplings do not obstruct the descent (a worry due to Benjamin Muntz, p.c.); what stands in the way of a \emph{Yukawa} employment of the colour determinant line is the hypercharge split between $u_R$ and $d_R$ (see footnote~\ref{ftnt:RH_split}), and what the $U(3)$ option would leave unexplained is the correlation of hypercharge with the weak representations \citep{Gomes_globalform}.} Charge quantisation (Section~\ref{par:charge}) follows from the tensorial construction of matter fields rather than from the topology of a compact group. The charge lattice persists even when the automorphism group is non-compact; a non-trivial point because the Lagrangian sees only the Lie algebra, so it doesn't prefer $U(1)$ over the additive group $\mathbb{R}$.

The structured fibres determine the available contraction tensors, the mass quadratic form, and its kernel; the choice of generating bundles fixes the charge lattice and the global form of the gauge group. But there are things that the geometry does \emph{not} determine. For instance, it is the gauge kinetic normalisation  that determines how $g$ and $g'$ enter; it is the Higgs potential that determines $v$ and $m_h$; and the Yukawa couplings and matrices remain free parameters. The reformulation redistributes explanatory labour, but it does not shrink the number of parameters in the Standard Model.

Thus I prefer the geometry-first picture on the grounds that it uses a sparser set of primitive posits for the theories to which it applies: no principal bundle, no independently postulated structure group. And while parsimony arguments rest on commitments that not everyone shares, the constraints of Section~\ref{sec:defense}, and the mechanisms derived under them, do not.

Neither does a second consideration in the reformulation's favour. For the classical Standard Model mechanisms treated here, the two formulations reproduce the same predictions, and one might conclude that nothing but taste is at stake. Feynman's Nobel Lecture says why that conclusion is too quick:
\begin{quote}
Theories of the known, which are described by different physical ideas may be equivalent in all their predictions and are hence scientifically indistinguishable. However, they are not psychologically identical when trying to move from that base into the unknown.
\end{quote}
His example, path integrals, equivalent to Schwinger's operator methods yet far more fertile, bear the point out; so does Minkowski's geometrical reformulation of special relativity, without which general relativity would scarcely have been possible. In this direction, \citet{Hunt_reform} argues  that reformulations are  valuable when they offer different primitive posits, new explanations, or a different way of solving problems. The geometry-first picture meets all three criteria.

There remains an objection from generality and explanatory reach. The construction formally accommodates exceptional groups,  but in the $E_8$ case it does so only by taking the gauge algebra and its bracket as primitive fibre structure. In the $E_8$ case, the group is no longer recovered from an independently motivated geometry of spaces where matter lives (Section~\ref{sec:E6}). 

 I concede that a single framework covering the widest possible class of theories is a genuine virtue of the principal-bundle picture. But I see no other source than this for a claim to priority for the principal bundle picture. \citet{Skinner_QFT} is unusually candid about this. After constructing principal bundles from frame bundles (assuming $G\simeq\mathsf{Aut}(V)$), he notes that ``the most common Lie groups that arise in physics are indeed matrix Lie groups [of that sort], so the two viewpoints are equivalent,'' but immediately qualifies: ``However, in some exotic theories \ldots exceptional Lie groups such as $E_8$ play an important role, so the fundamental picture is really that of principal bundles.'' Thus, for the Standard Model---and indeed for every empirically realised gauge theory---Skinner concedes the equivalence; the principal bundle's priority is justified solely by the wish to cover exotic theories. Moreover, the theories he calls ``common''---classical groups acting faithfully in fundamental representations, with matter built tensorially from a small set of fundamental spaces---are precisely those satisfying the geometry-first constraints. 

Indeed, practising physicists routinely treat the gauge group as determined by the matter content, and build representations tensorially, without formulating this as a principle. And although no standard textbook derives the Standard Model matter content from fundamental vector spaces, the working assumption satisfies exactly the condition in~\eqref{eq:G_VB}. The assumption is tacit, physicists are not always consistent about it, and BSM model-building with exceptional groups sometimes departs from it. But within the empirical domain, I take the geometry-first picture to articulate precisely what this practice is already tacitly committed to. And making a tacit assumption explicit is precisely the kind of work a reformulation should do.

\subsection*{Acknowledgements} 
I would like to especially thank  Aldo Riello, Benjamin Muntz,  Axel Maas, Silvester Borsboom, Benjamin Feintzeig, Doreen Fraser, Adam Koberiski, and Mark Hamilton for feedback and comments. I would also like to thank   David Tong, Oliver Pooley, David Wallace, Caspar Jacobs, Jim Weatherall, Jeremy Butterfield, and Eleanor March for many conversations on this topic. The authors gratefully acknowledge funding from the European Research Council, Grant 101088528
COGY
\appendix
\section*{APPENDIX}

\section{Principal and associated fibre bundles}\label{app:PFB}

A \emph{principal $G$-bundle} $(P,M,G)$ is a smooth manifold $P$ equipped with a free right action of a Lie group $G$, with projection $\pi:P\to P/G\simeq M$. Each fibre $\pi^{-1}(x)$ is diffeomorphic to $G$, the diffeomorphism being fixed by a choice of $p\in\pi^{-1}(x)$. A \emph{principal connection} $\varpi$ is a $\mathfrak{g}$-valued one-form on $P$ satisfying $\varpi(\xi^\#)=\xi$ (where $\xi^\#$ is the fundamental vector field generated by $\xi\in\mathfrak{g}$) and ${R_g}^*\varpi=\Ad_{g^{-1}}\varpi$. See \citet{Bleecker} or \citet{Gomes_elements} for details.

An \emph{associated vector bundle} is $E=P\times_\rho V$, where $\rho:G\to GL(V)$ is a representation and the equivalence relation identifies $(p,v)\sim(p\cdot g,\rho(g^{-1})v)$.\label{eq:AVB2} The connection $\varpi$ induces covariant derivatives on all associated bundles via horizontal lift:
\be\label{eq:PFB_cov}
\nabla_\mathbf{v}\kappa=\frac{d}{dt}[\gamma_h, \tau],
\ee
where $\kappa(x)=[\sigma(x),\tau(x)]$ and $\gamma_h$ is the horizontal lift of a curve tangent to $\mathbf{v}\in T_xM$.

A \emph{vector bundle} $(E,M,V)$ with covariant derivative $\nabla:\Gamma(E)\to\Gamma(T^*M\otimes E)$ satisfying the Leibniz rule $\nabla(f\kappa)=\d f\otimes\kappa+f\nabla\kappa$ determines a curvature $\Omega(X,Y)\kappa=(\nabla_X\nabla_Y-\nabla_Y\nabla_X-\nabla_{[X,Y]})\kappa$\label{def:curv}, an element of $\Omega^2(M,\mathsf{End}(E))$. The Yang--Mills Lagrangian is:
\be\label{eq:Lag_Omega}\mathcal{L}=\mathsf{Tr}(\Omega\wedge *\Omega).
\ee
The bundle of frames $L(E)$ is itself a principal $GL(V)$-bundle; enriching $V$ with further structure (inner product, volume form) restricts the admissible frames and selects a subgroup as the structure group.

\paragraph{Sourced Yang--Mills equations and vacuum gauge directions.}\label{app:sourced}
When $\rho:G\to U(V)$ forms the associated bundle $E=P\times_\rho V$, the sourced Yang--Mills equation reads $d_A^*F_A=J_H(A,\Phi)$.\label{eq:YM-sourced} If $\ker(\rho_*)\neq 0$, the current takes values only in $\ker(\rho_*)^\perp\subset\mathfrak{g}$, so the equation splits:
\be\label{eq:source}
\pi_{\ker(\rho_*)^\perp}(d_A^*F_A)=J_H(A,\Phi),
\qquad
\pi_{\ker(\rho_*)}(d_A^*F_A)=0.
\ee
Gauge directions that do not act on matter are automatically unsourced: these are vacuum gauge directions in the sense of Section~\ref{sec:slack}. If the kernel is only discrete ($\ker(\rho)\neq 0$ but $\ker(\rho_*)=0$), the local equations are unchanged; the difference concerns only global issues.

\section{Proofs}\label{app:proofs}

\begin{proof}[Proposition~\ref{prop:param}]
(i) The unit section $e_1:=\varphi/\|\varphi\|$ trivialises the line sub-bundle $L_1:=\mathrm{span}_{\mathbb{C}}\{e_1\}$, as $e_0$ does $L_0$. The orthogonal complements satisfy $\ell_1^\perp\simeq\det E\otimes L_1^*\simeq\det E\simeq\det E\otimes L_0^*\simeq\ell_0^\perp$; choose a unitary isomorphism $w:\ell_1^\perp\to\ell_0^\perp$. Then $u:=(e_1\mapsto e_0)\oplus w$ is a unitary automorphism of $E$ with $u\varphi=\|\varphi\|\,e_0$. (ii) Automorphisms are fibrewise unitary, so they preserve pointwise norms; the converse follows by applying (i) to both sections. (iii) Immediate from (i) and (ii); fibrewise, the stabiliser of a unit vector in $U(2)$ is the $U(1)$ acting on its orthogonal complement.
\end{proof}

\begin{proof}[Lemma~\ref{lem:mult_one}]
Each channel factorises as (colour)$\,\otimes\,$(weak)$\,\otimes\,$(hypercharge), so its invariant space is the tensor product of the factors' invariant spaces. Colour: trivial for the lepton channel; for the quark channels $\bar{\mathbf{3}}\otimes\mathbf{3}$ contains the trivial representation with multiplicity one. Weak: $\bar{\mathbf{2}}\otimes\mathbf{2}$ contains the trivial representation with multiplicity one (the Hermitian pairing); for the up-type channel $\bar{\mathbf{2}}\otimes\bar{\mathbf{2}}$ does so as well (the pairing through $\epsilon$). Hypercharge: the weights sum to zero in each channel (lepton: $+1+1-2$; down-type: $-\tfrac13+1-\tfrac23$; up-type: $-\tfrac13-1+\tfrac43$), so the resulting contraction is invariant, and each space is one-dimensional.
\end{proof}

\begin{proof}[Proposition~\ref{prop:central}]
$W$ carries a $G_W$-invariant Hermitian metric, and $G_W$ acts irreducibly (each matter fibre is a tensor product of irreducibles of distinct factors). Let $u\in N_{GL(W)}(G_W)$. First, $\langle u^{-1}\cdot,u^{-1}\cdot\rangle$ is $G_W$-invariant and positive-definite, hence equals $\lambda\langle\cdot,\cdot\rangle$ with $\lambda>0$ by irreducibility; so $u=\lambda^{1/2}u_0$ with $u_0$ unitary and normalising. Second, $\mathrm{Ad}_{u_0}$ is an automorphism of $G_W$; it preserves the simple ideals $\mathfrak{su}(3)$ and $\mathfrak{su}(2)$ of the image algebra (they are non-isomorphic) and fixes the $U(1)$ factor pointwise, since that factor acts on $W$ by scalars, which are central in $GL(W)$. Every automorphism of $SU(2)$ is inner. An outer automorphism of the $SU(3)$ factor would require $u_0$ to intertwine the colour representation on $W$ with its complex conjugate, which is inequivalent to it for every $W$ carrying non-trivial colour ($\mathbf{3}\not\simeq\bar{\mathbf{3}}$). Hence $\mathrm{Ad}_{u_0}$ is inner: $\mathrm{Ad}_{u_0}=\mathrm{Ad}_{\rho(g)}$ for some $g\in G$, and $\rho(g)^{-1}u_0\in C_{GL(W)}(G_W)=\mathbb{C}^\times$ by Schur's lemma. Thus $N_{GL(W)}(G_W)=\mathbb{C}^{\times}\!\cdot G_W$; and since every fibre in Table~\ref{tab:fermion-reps} carries non-zero hypercharge, the unit phases already lie in $G_W$, leaving $\mathbb{R}^{>0}$.
\end{proof}

\paragraph{Remark (the conjugate Higgs section).}
A complex-linear map extends factorwise over $\otimes_{\mathbb{C}}$; a conjugate-linear one does not, by the balancing computation of Section~\ref{sec:Yukawa2}. Conjugating every factor restores consistency: $J_H(v\otimes\ell)=J_2(v)\otimes J_L(\ell)$ is well defined, with $J_2, J_L$ the Riesz maps of the factors, and
\[
\widetilde{C}(v\otimes\ell)=(\epsilon^\#J_2\,v)\otimes J_L(\ell)=C(v)\otimes J_L(\ell).
\]
In a local unitary frame $(e_1,e_2)$ of $E^2$ with $\epsilon(e_1,e_2)=1$, and with $s$ a unit local section of $(E^1)^{\otimes3}$, a Higgs configuration $\varphi=(\varphi^1e_1+\varphi^2e_2)\otimes s$ goes to
\[
\widetilde{C}(\varphi)=\big(\bar\varphi^2\,e_1-\bar\varphi^1\,e_2\big)\otimes s^{*}:
\]
the familiar $\tilde\phi=i\sigma_2\bar\phi$, together with the replacement $s\mapsto s^{*}$ that component notation hides and that carries the hypercharge flip. Under $(g,z)\in SU(2)\times U(1)$, if $\varphi\mapsto (g\otimes z^{3})\,\varphi$ then $\widetilde{C}(\varphi)\mapsto (g\otimes z^{-3})\,\widetilde{C}(\varphi)$: the same $SU(2)$ transformation law, the opposite $U(1)$ charge: exactly what the up-type Yukawa channel requires.

%\bibliographystyle{apacite}
%\bibliography{references3}

\begin{thebibliography}{}

\bibitem [\protect \citeauthoryear {%
Baez%
\ \BBA {} Huerta%
}{%
Baez%
\ \BBA {} Huerta%
}{%
{\protect \APACyear {2010}}%
}]{%
BaezHuerta_GUT}
\APACinsertmetastar {%
BaezHuerta_GUT}%
\begin{APACrefauthors}%
Baez, J\BPBI C.%
\BCBT {}\ \BBA {} Huerta, J.%
\end{APACrefauthors}%
\unskip\
\newblock
\APACrefYearMonthDay{2010}{}{}.
\newblock
{\BBOQ}\APACrefatitle {{The algebra of grand unified theories}} {{The algebra
  of grand unified theories}}.{\BBCQ}
\newblock
\APACjournalVolNumPages{Bulletin of the American Mathematical
  Society}{47}{3}{483--552}.
\PrintBackRefs{\CurrentBib}

\bibitem [\protect \citeauthoryear {%
Baez%
\ \BBA {} Schwahn%
}{%
Baez%
\ \BBA {} Schwahn%
}{%
{\protect \APACyear {2026}}%
}]{%
BaezSchwahn_Jordan}
\APACinsertmetastar {%
BaezSchwahn_Jordan}%
\begin{APACrefauthors}%
Baez, J\BPBI C.%
\BCBT {}\ \BBA {} Schwahn, P.%
\end{APACrefauthors}%
\unskip\
\newblock
\APACrefYearMonthDay{2026}{}{}.
\newblock
\APACrefbtitle {{The standard model gauge group from the exceptional Jordan
  algebra}.} {{The standard model gauge group from the exceptional Jordan
  algebra}.}
\newblock
\APACrefnote{arXiv:2606.15235}
\PrintBackRefs{\CurrentBib}

\bibitem [\protect \citeauthoryear {%
Bhalla-Ladd%
, March%
\BCBL {}\ \BBA {} Weatherall%
}{%
Bhalla-Ladd%
\ \protect \BOthers {.}}{%
{\protect \APACyear {2026}}%
}]{%
BhallaLadd_gluon}
\APACinsertmetastar {%
BhallaLadd_gluon}%
\begin{APACrefauthors}%
Bhalla-Ladd, I.%
, March, E.%
\BCBL {}\ \BBA {} Weatherall, J\BPBI O.%
\end{APACrefauthors}%
\unskip\
\newblock
\APACrefYearMonthDay{2026}{}{}.
\newblock
\APACrefbtitle {{Ceci n'est pas un gluon}.} {{Ceci n'est pas un gluon}.}
\newblock
\APACrefnote{arXiv:2603.19518}
\PrintBackRefs{\CurrentBib}

\bibitem [\protect \citeauthoryear {%
Bleecker%
}{%
Bleecker%
}{%
{\protect \APACyear {1981}}%
}]{%
Bleecker}
\APACinsertmetastar {%
Bleecker}%
\begin{APACrefauthors}%
Bleecker, D.%
\end{APACrefauthors}%
\unskip\
\newblock
\APACrefYear{1981}.
\newblock
\APACrefbtitle {{Gauge Theory and Variational Principles}} {{Gauge Theory and
  Variational Principles}}.
\newblock
\APACaddressPublisher{}{Dover Publications}.
\PrintBackRefs{\CurrentBib}

\bibitem [\protect \citeauthoryear {%
Boyle%
}{%
Boyle%
}{%
{\protect \APACyear {2026}}%
}]{%
Boyle_triality}
\APACinsertmetastar {%
Boyle_triality}%
\begin{APACrefauthors}%
Boyle, L.%
\end{APACrefauthors}%
\unskip\
\newblock
\APACrefYearMonthDay{2026}{}{}.
\newblock
{\BBOQ}\APACrefatitle {{The standard model, the exceptional Jordan algebra, and
  triality}} {{The standard model, the exceptional Jordan algebra, and
  triality}}.{\BBCQ}
\newblock
\APACjournalVolNumPages{Journal of Mathematical Physics}{67}{7}{071701}.
\PrintBackRefs{\CurrentBib}

\bibitem [\protect \citeauthoryear {%
Boyle%
\ \BBA {} Farnsworth%
}{%
Boyle%
\ \BBA {} Farnsworth%
}{%
{\protect \APACyear {2014}}%
}]{%
BoyleFarnsworth_NJP}
\APACinsertmetastar {%
BoyleFarnsworth_NJP}%
\begin{APACrefauthors}%
Boyle, L.%
\BCBT {}\ \BBA {} Farnsworth, S.%
\end{APACrefauthors}%
\unskip\
\newblock
\APACrefYearMonthDay{2014}{}{}.
\newblock
{\BBOQ}\APACrefatitle {{Non-commutative geometry, non-associative geometry and
  the standard model of particle physics}} {{Non-commutative geometry,
  non-associative geometry and the standard model of particle physics}}.{\BBCQ}
\newblock
\APACjournalVolNumPages{New Journal of Physics}{16}{}{123027}.
\PrintBackRefs{\CurrentBib}

\bibitem [\protect \citeauthoryear {%
Brading%
\ \BBA {} Castellani%
}{%
Brading%
\ \BBA {} Castellani%
}{%
{\protect \APACyear {2003}}%
}]{%
Brading_Castellani}
\APACinsertmetastar {%
Brading_Castellani}%
\begin{APACrefauthors}%
Brading, K.%
\BCBT {}\ \BBA {} Castellani, E.%
\end{APACrefauthors}%
\unskip\
\newblock
\APACrefYear{2003}.
\newblock
\APACrefbtitle {{Symmetries in Physics: Philosophical Reflections}} {{Symmetries in physics: philosophical reflections}}.
\newblock
\APACaddressPublisher{Cambridge}{Cambridge University Press}.
\PrintBackRefs{\CurrentBib}

\bibitem [\protect \citeauthoryear {%
Brown%
}{%
Brown%
}{%
{\protect \APACyear {2005}}%
}]{%
Brown_PR}
\APACinsertmetastar {%
Brown_PR}%
\begin{APACrefauthors}%
Brown, H\BPBI R.%
\end{APACrefauthors}%
\unskip\
\newblock
\APACrefYear{2005}.
\newblock
\APACrefbtitle {{Physical Relativity: Space-Time Structure from a Dynamical Perspective}} {{Physical relativity: space-time structure from a dynamical perspective}}.
\newblock
\APACaddressPublisher{Oxford}{Oxford University Press}.
\PrintBackRefs{\CurrentBib}

\bibitem [\protect \citeauthoryear {%
Chamseddine%
\ \BBA {} Connes%
}{%
Chamseddine%
\ \BBA {} Connes%
}{%
{\protect \APACyear {1997}}%
}]{%
ChamseddineConnes_spectral}
\APACinsertmetastar {%
ChamseddineConnes_spectral}%
\begin{APACrefauthors}%
Chamseddine, A\BPBI H.%
\BCBT {}\ \BBA {} Connes, A.%
\end{APACrefauthors}%
\unskip\
\newblock
\APACrefYearMonthDay{1997}{}{}.
\newblock
{\BBOQ}\APACrefatitle {{The spectral action principle}} {{The spectral action
  principle}}.{\BBCQ}
\newblock
\APACjournalVolNumPages{Communications in Mathematical
  Physics}{186}{}{731--750}.
\PrintBackRefs{\CurrentBib}

\bibitem [\protect \citeauthoryear {%
Chamseddine%
\ \BBA {} Connes%
}{%
Chamseddine%
\ \BBA {} Connes%
}{%
{\protect \APACyear {2012}}%
}]{%
ChamseddineConnes_resilience}
\APACinsertmetastar {%
ChamseddineConnes_resilience}%
\begin{APACrefauthors}%
Chamseddine, A\BPBI H.%
\BCBT {}\ \BBA {} Connes, A.%
\end{APACrefauthors}%
\unskip\
\newblock
\APACrefYearMonthDay{2012}{}{}.
\newblock
{\BBOQ}\APACrefatitle {{Resilience of the spectral standard model}}
  {{Resilience of the spectral standard model}}.{\BBCQ}
\newblock
\APACjournalVolNumPages{Journal of High Energy Physics}{2012}{9}{104}.
\PrintBackRefs{\CurrentBib}

\bibitem [\protect \citeauthoryear {%
Chamseddine%
, Connes%
\BCBL {}\ \BBA {} Marcolli%
}{%
Chamseddine%
\ \protect \BOthers {.}}{%
{\protect \APACyear {2007}}%
}]{%
CCM_gravity}
\APACinsertmetastar {%
CCM_gravity}%
\begin{APACrefauthors}%
Chamseddine, A\BPBI H.%
, Connes, A.%
\BCBL {}\ \BBA {} Marcolli, M.%
\end{APACrefauthors}%
\unskip\
\newblock
\APACrefYearMonthDay{2007}{}{}.
\newblock
{\BBOQ}\APACrefatitle {{Gravity and the standard model with neutrino mixing}}
  {{Gravity and the standard model with neutrino mixing}}.{\BBCQ}
\newblock
\APACjournalVolNumPages{Advances in Theoretical and Mathematical
  Physics}{11}{6}{991--1089}.
\PrintBackRefs{\CurrentBib}

\bibitem [\protect \citeauthoryear {%
Coleman%
}{%
Coleman%
}{%
{\protect \APACyear {1985}}%
}]{%
Coleman_Aspects}
\APACinsertmetastar {%
Coleman_Aspects}%
\begin{APACrefauthors}%
Coleman, S.%
\end{APACrefauthors}%
\unskip\
\newblock
\APACrefYear{1985}.
\newblock
\APACrefbtitle {{Aspects of Symmetry: Selected Erice Lectures}} {{Aspects of
  Symmetry: Selected Erice Lectures}}.
\newblock
\APACaddressPublisher{Cambridge}{Cambridge University Press}.
\PrintBackRefs{\CurrentBib}

\bibitem [\protect \citeauthoryear {%
Elitzur%
}{%
Elitzur%
}{%
{\protect \APACyear {1975}}%
}]{%
Elitzur1975}
\APACinsertmetastar {%
Elitzur1975}%
\begin{APACrefauthors}%
Elitzur, S.%
\end{APACrefauthors}%
\unskip\
\newblock
\APACrefYearMonthDay{1975}{}{}.
\newblock
{\BBOQ}\APACrefatitle {Impossibility of spontaneously breaking local symmetries} {Impossibility of spontaneously breaking local symmetries}.{\BBCQ}
\newblock
\APACjournalVolNumPages{Physical Review D}{12}{12}{3978--3982}.
\newblock
\begin{APACrefDOI} \doi{10.1103/PhysRevD.12.3978} \end{APACrefDOI}
\PrintBackRefs{\CurrentBib}

\bibitem [\protect \citeauthoryear {%
Furey%
}{%
Furey%
}{%
{\protect \APACyear {2015}}%
}]{%
Furey_charge}
\APACinsertmetastar {%
Furey_charge}%
\begin{APACrefauthors}%
Furey, C.%
\end{APACrefauthors}%
\unskip\
\newblock
\APACrefYearMonthDay{2015}{}{}.
\newblock
{\BBOQ}\APACrefatitle {{Charge quantization from a number operator}} {{Charge
  quantization from a number operator}}.{\BBCQ}
\newblock
\APACjournalVolNumPages{Physics Letters B}{742}{}{195--199}.
\PrintBackRefs{\CurrentBib}

\bibitem [\protect \citeauthoryear {%
Furey%
}{%
Furey%
}{%
{\protect \APACyear {2018}}%
}]{%
Furey_ladder}
\APACinsertmetastar {%
Furey_ladder}%
\begin{APACrefauthors}%
Furey, C.%
\end{APACrefauthors}%
\unskip\
\newblock
\APACrefYearMonthDay{2018}{}{}.
\newblock
{\BBOQ}\APACrefatitle {{$SU(3)_C\times SU(2)_L\times U(1)_Y(\times U(1)_X)$ as
  a symmetry of division algebraic ladder operators}} {{$SU(3)_C\times
  SU(2)_L\times U(1)_Y(\times U(1)_X)$ as a symmetry of division algebraic
  ladder operators}}.{\BBCQ}
\newblock
\APACjournalVolNumPages{The European Physical Journal C}{78}{}{375}.
\PrintBackRefs{\CurrentBib}

\bibitem [\protect \citeauthoryear {%
Geroch%
}{%
Geroch%
}{%
{\protect \APACyear {1972}}%
}]{%
Geroch1972}
\APACinsertmetastar {%
Geroch1972}%
\begin{APACrefauthors}%
Geroch, R.%
\end{APACrefauthors}%
\unskip\
\newblock
\APACrefYearMonthDay{1972}{dec}{}.
\newblock
{\BBOQ}\APACrefatitle {Einstein algebras} {Einstein algebras}.{\BBCQ}
\newblock
\APACjournalVolNumPages{Communications in Mathematical
  Physics}{26}{4}{271--275}.
\newblock
\begin{APACrefDOI} \doi{10.1007/bf01645521} \end{APACrefDOI}
\PrintBackRefs{\CurrentBib}

\bibitem [\protect \citeauthoryear {%
Gomes%
}{%
Gomes%
}{%
{\protect \APACyear {2024}}%
{\protect \APACexlab {{\protect \BCnt {1}}}}%
}]{%
Gomes_internal}
\APACinsertmetastar {%
Gomes_internal}%
\begin{APACrefauthors}%
Gomes, H.%
\end{APACrefauthors}%
\unskip\
\newblock
\APACrefYearMonthDay{2024{\protect \BCnt {1}}}{{\APACmonth{10}}}{}.
\newblock
{\BBOQ}\APACrefatitle {{Gauge Theory Without Principal Fiber Bundles}} {{Gauge
  Theory Without Principal Fiber Bundles}}.{\BBCQ}
\newblock
\APACjournalVolNumPages{Philosophy of Science}{}{}{1--17}.
\newblock
\begin{APACrefDOI} \doi{10.1017/psa.2024.49} \end{APACrefDOI}
\PrintBackRefs{\CurrentBib}

\bibitem [\protect \citeauthoryear {%
Gomes%
}{%
Gomes%
}{%
{\protect \APACyear {2024}}%
{\protect \APACexlab {{\protect \BCnt {2}}}}%
}]{%
Gomes_elements}
\APACinsertmetastar {%
Gomes_elements}%
\begin{APACrefauthors}%
Gomes, H.%
\end{APACrefauthors}%
\unskip\
\newblock
\APACrefYear{2024{\protect \BCnt {2}}}.
\newblock
\APACrefbtitle {{Gauge Theory and the Geometrisation of Physics}} {{Gauge theory and the geometrisation of physics}}.
\newblock
\APACaddressPublisher{}{Cambridge University Press}.
\PrintBackRefs{\CurrentBib}

\bibitem [\protect \citeauthoryear {%
Gomes%
}{%
Gomes%
}{%
{\protect \APACyear {2025}}%
{\protect \APACexlab {{\protect \BCnt {1}}}}}]{%
Gomes_AB}
\APACinsertmetastar {%
Gomes_AB}%
\begin{APACrefauthors}%
Gomes, H.%
\end{APACrefauthors}%
\unskip\
\newblock
\APACrefYearMonthDay{2025{\protect \BCnt {1}}}{}{}.
\newblock
{\BBOQ}\APACrefatitle {{The Aharonov-Bohm effect: fact and reality}} {{The
  Aharonov-Bohm effect: fact and reality}}.{\BBCQ}
\newblock
\APACjournalVolNumPages{Philosophy of Physics}{}{}{}.
\PrintBackRefs{\CurrentBib}

\bibitem [\protect \citeauthoryear {%
Gomes%
}{%
Gomes%
}{%
{\protect \APACyear {2025}}%
{\protect \APACexlab {{\protect \BCnt {2}}}}}]{%
Rep_conv}
\APACinsertmetastar {%
Rep_conv}%
\begin{APACrefauthors}%
Gomes, H.%
\end{APACrefauthors}%
\unskip\
\newblock
\APACrefYearMonthDay{2025{\protect \BCnt {2}}}{}{}.
\newblock
{\BBOQ}\APACrefatitle {{Representational Schemes for theories with symmetries}}
  {{Representational Schemes for theories with symmetries}}.{\BBCQ}
\newblock
\APACjournalVolNumPages{Synthese}{}{}{}.
\PrintBackRefs{\CurrentBib}

\bibitem [\protect \citeauthoryear {%
Gomes%
}{%
Gomes%
}{%
{\protect \APACyear {2026}}%
{\protect \APACexlab {{\protect \BCnt {1}}}}%
}]{%
Gomes_nonequiv}
\APACinsertmetastar {%
Gomes_nonequiv}%
\begin{APACrefauthors}%
Gomes, H.%
\end{APACrefauthors}%
\unskip\
\newblock
\APACrefYearMonthDay{2026{\protect \BCnt {1}}}{}{}.
\newblock
{\BBOQ}\APACrefatitle {{The geometry-first formulation of gauge theory is not equivalent to the symmetry-first one}} {{The geometry-first formulation of gauge theory is not equivalent to the symmetry-first one}}.{\BBCQ}
\newblock
\APACrefnote{arXiv:2607.24901}
\PrintBackRefs{\CurrentBib}

\bibitem [\protect \citeauthoryear {%
Gomes%
}{%
Gomes%
}{%
{\protect \APACyear {2026}}%
{\protect \APACexlab {{\protect \BCnt {2}}}}%
}]{%
Gomes_globalform}
\APACinsertmetastar {%
Gomes_globalform}%
\begin{APACrefauthors}%
Gomes, H.%
\end{APACrefauthors}%
\unskip\
\newblock
\APACrefYearMonthDay{2026{\protect \BCnt {2}}}{}{}.
\newblock
{\BBOQ}\APACrefatitle {{Geometry first and the global form of the Standard Model gauge group}} {{Geometry first and the global form of the Standard Model gauge group}}.{\BBCQ}
\newblock
\APACjournalVolNumPages{Manuscript}{}{}{}.
\PrintBackRefs{\CurrentBib}

\bibitem [\protect \citeauthoryear {%
Gomes%
\ \BBA {} Butterfield%
}{%
Gomes%
\ \BBA {} Butterfield%
}{%
{\protect \APACyear {2023}}%
}]{%
GomesButterfield_hole2}
\APACinsertmetastar {%
GomesButterfield_hole2}%
\begin{APACrefauthors}%
Gomes, H.%
\BCBT {}\ \BBA {} Butterfield, J.%
\end{APACrefauthors}%
\unskip\
\newblock
\APACrefYearMonthDay{2023}{{\APACmonth{06}}}{}.
\newblock
{\BBOQ}\APACrefatitle {{The Hole Argument and Beyond: Part II: Treating
  Non-isomorphic Spacetimes}} {{The Hole Argument and Beyond: Part II: Treating
  Non-isomorphic Spacetimes}}.{\BBCQ}
\newblock
\APACjournalVolNumPages{Journal of Physics: Conference
  Series}{2533}{1}{012003}.
\newblock
\begin{APACrefURL} \url{https://dx.doi.org/10.1088/1742-6596/2533/1/012003}
  \end{APACrefURL}
\newblock
\begin{APACrefDOI} \doi{10.1088/1742-6596/2533/1/012003} \end{APACrefDOI}
\PrintBackRefs{\CurrentBib}

\bibitem[\protect\citeauthoryear{Gomes \BIBand\ Muntz}{Gomes \BIBand\ Muntz}{{\APACyear{2026}}}]{Anomalies_new}
\APACinsertmetastar{Anomalies_new}%
Gomes, H.\BCBT\ \BBA\ Muntz, B.
\unskip\
\newblock
\APACrefYearMonthDay{2026}{}{}.
\newblock
{\BBOQ}\APACrefatitle{Local Anomalies and Hypercharge in the Geometry-First Formulation of the Standard Model}{Local anomalies and hypercharge in the geometry-first formulation of the Standard Model}.{\BBCQ}
\newblock
\APACjournalVolNumPages{Manuscript in preparation}{}{}{}.
\PrintBackRefs{\CurrentBib}

\bibitem [\protect \citeauthoryear {%
Gomes%
\ \BBA {} Riello%
}{%
Gomes%
\ \BBA {} Riello%
}{%
{\protect \APACyear {2024}}%
}]{%
GomesRiello_theta}
\APACinsertmetastar {%
GomesRiello_theta}%
\begin{APACrefauthors}%
Gomes, H.%
\BCBT {}\ \BBA {} Riello, A.%
\end{APACrefauthors}%
\unskip\
\newblock
\APACrefYearMonthDay{2024}{}{}.
\newblock
{\BBOQ}\APACrefatitle {Eliminativism and the {QCD} $\theta_{\text{YM}}$-Term: What Gauge Transformations Cannot Do} {Eliminativism and the {QCD} $\theta_{\text{YM}}$-term: what gauge transformations cannot do}.{\BBCQ}
\newblock
\APACjournalVolNumPages{Foundations of Physics}{54}{}{40}.
\newblock
\begin{APACrefDOI} \doi{10.1007/s10701-024-00759-5} \end{APACrefDOI}
\PrintBackRefs{\CurrentBib}

\bibitem [\protect \citeauthoryear {%
Hamilton%
}{%
Hamilton%
}{%
{\protect \APACyear {2017}}%
}]{%
Hamilton_book}
\APACinsertmetastar {%
Hamilton_book}%
\begin{APACrefauthors}%
Hamilton, M.%
\end{APACrefauthors}%
\unskip\
\newblock
\APACrefYear{2017}.
\newblock
\APACrefbtitle {{Mathematical Gauge Theory}} {{Mathematical Gauge Theory}}.
\newblock
\APACaddressPublisher{}{Springer International Publishing}.
\newblock
\begin{APACrefDOI} \doi{10.1007/978-3-319-68439-0} \end{APACrefDOI}
\PrintBackRefs{\CurrentBib}

\bibitem [\protect \citeauthoryear {%
Hunt%
}{%
Hunt%
}{%
{\protect \APACyear {2025}}%
}]{%
Hunt_reform}
\APACinsertmetastar {%
Hunt_reform}%
\begin{APACrefauthors}%
Hunt, J.%
\end{APACrefauthors}%
\unskip\
\newblock
\APACrefYearMonthDay{2025}{}{}.
\newblock
{\BBOQ}\APACrefatitle {{On the Value of Reformulating}} {{On the Value of
  Reformulating}}.{\BBCQ}
\newblock
\APACjournalVolNumPages{Journal of Philosophy}{}{}{}.
\PrintBackRefs{\CurrentBib}

\bibitem [\protect \citeauthoryear {%
Jacobs%
}{%
Jacobs%
}{%
{\protect \APACyear {2023}}%
}]{%
Jacobs_PFB}
\APACinsertmetastar {%
Jacobs_PFB}%
\begin{APACrefauthors}%
Jacobs, C.%
\end{APACrefauthors}%
\unskip\
\newblock
\APACrefYearMonthDay{2023}{}{}.
\newblock
{\BBOQ}\APACrefatitle {The metaphysics of fibre bundles} {The metaphysics of
  fibre bundles}.{\BBCQ}
\newblock
\APACjournalVolNumPages{Studies in History and Philosophy of
  Science}{97}{}{34-43}.
\newblock
\begin{APACrefURL}
  \url{https://www.sciencedirect.com/science/article/pii/S0039368122001777}
  \end{APACrefURL}
\newblock
\begin{APACrefDOI} \doi{https://doi.org/10.1016/j.shpsa.2022.11.010}
  \end{APACrefDOI}
\PrintBackRefs{\CurrentBib}

\bibitem [\protect \citeauthoryear {%
Kabel%
\ \protect \BOthers {.}}{%
Kabel%
\ \protect \BOthers {.}}{%
{\protect \APACyear {2025}}%
}]{%
Kabel2025}
\APACinsertmetastar {%
Kabel2025}%
\begin{APACrefauthors}%
Kabel, V.%
, de~la Hamette, A\BHBI C.%
, Apadula, L.%
, Cepollaro, C.%
, Gomes, H.%
, Butterfield, J.%
\BCBL {}\ \BBA {} Brukner, C.%
\end{APACrefauthors}%
\unskip\
\newblock
\APACrefYearMonthDay{2025}{{\APACmonth{04}}}{}.
\newblock
{\BBOQ}\APACrefatitle {Quantum coordinates, localisation of events, and the
  quantum hole argument} {Quantum coordinates, localisation of events, and the
  quantum hole argument}.{\BBCQ}
\newblock
\APACjournalVolNumPages{Communications Physics}{8}{1}{}.
\newblock
\begin{APACrefDOI} \doi{10.1038/s42005-025-02084-3} \end{APACrefDOI}
\PrintBackRefs{\CurrentBib}

\bibitem [\protect \citeauthoryear {%
Krasnov%
}{%
Krasnov%
}{%
{\protect \APACyear {2021}}%
}]{%
Krasnov_SO9}
\APACinsertmetastar {%
Krasnov_SO9}%
\begin{APACrefauthors}%
Krasnov, K.%
\end{APACrefauthors}%
\unskip\
\newblock
\APACrefYearMonthDay{2021}{}{}.
\newblock
{\BBOQ}\APACrefatitle {{$SO(9)$ characterization of the standard model gauge
  group}} {{$SO(9)$ characterization of the standard model gauge
  group}}.{\BBCQ}
\newblock
\APACjournalVolNumPages{Journal of Mathematical Physics}{62}{}{021703}.
\PrintBackRefs{\CurrentBib}

\bibitem [\protect \citeauthoryear {%
Lange%
}{%
Lange%
}{%
{\protect \APACyear {2007}}%
}]{%
Lange_metalaws}
\APACinsertmetastar {%
Lange_metalaws}%
\begin{APACrefauthors}%
Lange, M.%
\end{APACrefauthors}%
\unskip\
\newblock
\APACrefYearMonthDay{2007}{}{}.
\newblock
{\BBOQ}\APACrefatitle {Laws and Meta-Laws of Nature: Conservation Laws and Symmetries} {Laws and meta-laws of nature: conservation laws and symmetries}.{\BBCQ}
\newblock
\APACjournalVolNumPages{Studies in History and Philosophy of Modern Physics}{38}{3}{457--481}.
\newblock
\begin{APACrefDOI} \doi{10.1016/j.shpsb.2006.08.003} \end{APACrefDOI}
\PrintBackRefs{\CurrentBib}

\bibitem [\protect \citeauthoryear {%
Michor%
}{%
Michor%
}{%
{\protect \APACyear {2008}}%
}]{%
Michor2008}
\APACinsertmetastar {%
Michor2008}%
\begin{APACrefauthors}%
Michor, P\BPBI W.%
\end{APACrefauthors}%
\unskip\
\newblock
\APACrefYear{2008}.
\newblock
\APACrefbtitle {Topics in differential geometry} {Topics in differential
  geometry}\ (\BNUM\ volume 93).
\newblock
\APACaddressPublisher{Providence, Rhode Island}{American Mathematical Society}.
\newblock
\APACrefnote{Includes bibliographical references (pages 479-488) and index.
  Description based on print version record.}
\PrintBackRefs{\CurrentBib}

\bibitem [\protect \citeauthoryear {%
Skinner%
}{%
Skinner%
}{%
{\protect \APACyear {2017}}%
}]{%
Skinner_QFT}
\APACinsertmetastar {%
Skinner_QFT}%
\begin{APACrefauthors}%
Skinner, D.%
\end{APACrefauthors}%
\unskip\
\newblock
\APACrefYear{2017}.
\newblock
\APACrefbtitle {{Quantum Field Theory II}} {{Quantum Field Theory II}}.
\newblock
\APACaddressPublisher{}{University of Cambridge lecture notes}.
\PrintBackRefs{\CurrentBib}

\bibitem [\protect \citeauthoryear {%
Susskind%
}{%
Susskind%
}{%
{\protect \APACyear {2026}}%
}]{%
Susskind2026}
\APACinsertmetastar {%
Susskind2026}%
\begin{APACrefauthors}%
Susskind, L.%
\end{APACrefauthors}%
\unskip\
\newblock
\APACrefYearMonthDay{2026}{}{}.
\newblock
{\BBOQ}\APACrefatitle {{Is Time Reversal in de Sitter Space a Spontaneously
  Broken Gauge Symmetry?}} {{Is Time Reversal in de Sitter Space a
  Spontaneously Broken Gauge Symmetry?}}.{\BBCQ}
\newblock
\APACjournalVolNumPages{arXiv preprint arXiv:2603.12434}{}{}{}.
\PrintBackRefs{\CurrentBib}

\bibitem [\protect \citeauthoryear {%
Todorov%
\ \BBA {} Dubois-Violette%
}{%
Todorov%
\ \BBA {} Dubois-Violette%
}{%
{\protect \APACyear {2018}}%
}]{%
DVTodorov_Jordan}
\APACinsertmetastar {%
DVTodorov_Jordan}%
\begin{APACrefauthors}%
Todorov, I.%
\BCBT {}\ \BBA {} Dubois-Violette, M.%
\end{APACrefauthors}%
\unskip\
\newblock
\APACrefYearMonthDay{2018}{}{}.
\newblock
{\BBOQ}\APACrefatitle {{Deducing the symmetry of the standard model from the
  automorphism and structure groups of the exceptional Jordan algebra}}
  {{Deducing the symmetry of the standard model from the automorphism and
  structure groups of the exceptional Jordan algebra}}.{\BBCQ}
\newblock
\APACjournalVolNumPages{International Journal of Modern Physics
  A}{33}{20}{1850118}.
\PrintBackRefs{\CurrentBib}

\bibitem [\protect \citeauthoryear {%
Tong%
}{%
Tong%
}{%
{\protect \APACyear {2017}}%
}]{%
Tong_line}
\APACinsertmetastar {%
Tong_line}%
\begin{APACrefauthors}%
Tong, D.%
\end{APACrefauthors}%
\unskip\
\newblock
\APACrefYearMonthDay{2017}{}{}.
\newblock
{\BBOQ}\APACrefatitle {{Line operators in the standard model}} {{Line operators
  in the standard model}}.{\BBCQ}
\newblock
\APACjournalVolNumPages{Journal of High Energy Physics}{2017}{7}{104}.
\PrintBackRefs{\CurrentBib}

\bibitem [\protect \citeauthoryear {%
Tong%
}{%
Tong%
}{%
{\protect \APACyear {2025}}%
}]{%
Tong_SM}
\APACinsertmetastar {%
Tong_SM}%
\begin{APACrefauthors}%
Tong, D.%
\end{APACrefauthors}%
\unskip\
\newblock
\APACrefYear{2025}.
\newblock
\APACrefbtitle {{The Standard Model}} {{The Standard Model}}.
\newblock
\APACaddressPublisher{}{Cambridge University Press}.
\PrintBackRefs{\CurrentBib}

\bibitem [\protect \citeauthoryear {%
Weatherall%
}{%
Weatherall%
}{%
{\protect \APACyear {2016}}%
}]{%
Weatherall2016_YMGR}
\APACinsertmetastar {%
Weatherall2016_YMGR}%
\begin{APACrefauthors}%
Weatherall, J.%
\end{APACrefauthors}%
\unskip\
\newblock
\APACrefYearMonthDay{2016}{}{}.
\newblock
{\BBOQ}\APACrefatitle {{Fiber bundles, Yang--Mills theory, and general
  relativity}} {{Fiber bundles, Yang--Mills theory, and general
  relativity}}.{\BBCQ}
\newblock
\APACjournalVolNumPages{Synthese}{193}{8}{2389--2425}.
\newblock
\APACrefnote{\url{http://philsci-archive.pitt.edu/11481/}}
\PrintBackRefs{\CurrentBib}

\end{thebibliography}

\end{document}